\documentclass[a4paper,11pt]{article}
\pdfoutput=1

\usepackage{jcappub}
\usepackage{amsmath}
\usepackage{amsfonts}
\usepackage{amssymb}
\usepackage{mathtools}
\usepackage{graphics}
\usepackage{tabularx}
\usepackage[export]{adjustbox}
\usepackage{multirow}
\usepackage{citesort}
\usepackage{graphicx}
\usepackage{url}
\usepackage{soul}
\usepackage{physics}
\usepackage{bm}
\usepackage[dvipsnames]{xcolor}
\usepackage[utf8]{inputenc}
\usepackage[normalem]{ulem}
\usepackage{hyperref}

\usepackage{tikz,xcolor,hyperref}

\definecolor{lime}{HTML}{A6CE39}
\DeclareRobustCommand{\orcidicon}{\hspace{-1mm}
 \begin{tikzpicture}
 \draw[lime, fill=lime] (0,0) 
 circle [radius=0.16] 
 node[white] {{\fontfamily{qag}\selectfont \tiny \,ID}};
 \draw[white, fill=white] (-0.0525,0.095) 
 circle [radius=0.007];
 \end{tikzpicture}
 \hspace{-3mm}
}

\foreach \x in {A, ..., Z}{\expandafter\xdef\csname orcid\x\endcsname{\noexpand\href{https://orcid.org/\csname orcidauthor\x\endcsname}
 {\noexpand\orcidicon}}
}

\title{Neutrino quantum kinetics in two spatial dimensions}

\author[a]{Marie Cornelius\orcidA{},}
\author[a]{Shashank Shalgar\orcidB{}}
\author[a]{and Irene Tamborra\orcidC{}}

\affiliation[a]{Niels Bohr International Academy and DARK, Niels Bohr Institute, University of Copenhagen, Blegdamsvej 17, 2100, Copenhagen, Denmark}

\emailAdd{marie.cornelius@nbi.ku.dk}
\emailAdd{shashank.shalgar@nbi.ku.dk}
\emailAdd{tamborra@nbi.ku.dk}

\abstract{Our understanding of neutrino flavor conversion in the innermost regions of core-collapse supernovae and neutron star mergers is mostly limited to spherically symmetric configurations that facilitate the numerical solution of the quantum kinetic equations. In this paper, we simulate neutrino quantum kinetics within a $(2+1+1)$ dimensional setup: we model the flavor evolution during neutrino decoupling from matter in two spatial dimensions, one neutrino momentum variable, and time; taking into account non-forward neutral current and charged current collisions of neutrinos with the matter background, as well as neutrino advection. In order to mimic fluctuations in the neutrino emission and matter background, and explore their effect on the flavor evolution, we introduce perturbations in the collision term as well as in the vacuum term of the Hamiltonian. Because of such perturbations, the initial symmetry of the neutrino field across the simulation annulus is broken and flavor conversion is qualitatively affected, with regions of larger flavor conversion alternating across the simulation annulus. In addition, neutrino advection is responsible for spreading flavor waves across neighboring spatial regions. Although based on a simplified setup, our findings highlight the importance of modeling neutrino quantum kinetics in multi-dimensions to assess the impact of neutrinos on the physics of compact astrophysical sources and nucleosynthesis.
}

\begin{document}
\maketitle

\section{Introduction}
Neutrinos play a fundamental role in core-collapse supernovae and neutron star mergers, despite their weakly interacting nature~\cite{Burrows:2020qrp,Mirizzi:2015eza,Mezzacappa:2020oyq,Janka:2016fox,Muller:2020ard}.
In the core of these sources, the neutrino number density is so large that the coherent forward scattering of neutrinos off each other significantly affects the flavor evolution and makes the neutrino equations of motion non-linear~\cite{Pantaleone:1994ns,Mirizzi:2015eza, Duan:2010bg, Chakraborty:2016yeg, Tamborra:2020cul,Richers:2022zug}.

In order to model the rich flavor phenomenology resulting from flavor conversion in neutrino-dense astrophysical sources and grasp its consequences on the source physics and multi-messenger observables~\cite{Ehring:2023lcd,Ehring:2023abs,Nagakura:2023mhr,Wu:2017drk,George:2020veu,Just:2022flt,Fernandez:2022yyv,Li:2021vqj}, a quantum-kinetic treatment of the Boltzmann neutrino transport is necessary~\cite{Tamborra:2020cul,Richers:2022zug,Mezzacappa:2020oyq,Nagakura:2022qko,Foucart:2022bth,Richers:2019grc}. This is a computationally challenging task due to the fact that the characteristic quantities entering the neutrino quantum kinetic equations (QKEs) change over several orders of magnitude; additionally the QKEs should be solved in seven dimensions (i.e., time, three spatial coordinates and three neutrino momentum coordinates). As a consequence, this problem can only be tackled relying on simplified setups, imposing symmetry assumptions on the system to reduce its dimensionality~\cite{Shalgar:2022rjj, Shalgar:2022lvv,Nagakura:2022xwe,Cornelius:2023eop,Shalgar:2019qwg,Xiong:2024tac,Martin:2021xyl,Xiong:2023vcm,Wu:2021uvt,Richers:2021nbx,Shalgar:2024gjt,Martin:2019gxb,Richers:2022bkd,Richers:2021xtf}.

Crossings in the electron lepton number (ELN) angular distribution of neutrinos can develop as neutrinos decouple from matter, induce flavor instabilities~\cite{Izaguirre:2016gsx,Chakraborty:2016lct,Morinaga:2021vmc,Fiorillo:2024bzm}, and eventually fast flavor conversion~\cite{Padilla-Gay:2021haz,Fiorillo:2023mze,Fiorillo:2023hlk}. The latter is characterized by a frequency directly proportional to the number density of neutrinos and could take place in the absence of neutrino mass difference. The advection of neutrinos may be responsible for smearing the ELN structures responsible for flavor conversion, unless such structures are self-sustained in time~\cite{Shalgar:2019qwg,Padilla-Gay:2020uxa}. In addition, non-forward collisions of neutrinos with the background medium can dynamically affect the flavor evolution~\cite{Shalgar:2020wcx,Hansen:2020vgm,Martin:2021xyl,Sigl:2021tmj,Sasaki:2021zld} or lead to the development of collisional flavor instabilities~\cite{Johns:2022yqy,Padilla-Gay:2022wck,Lin:2022dek,Xiong:2022vsy,Liu:2023vtz,Fiorillo:2023ajs}, whose relevance in the core of astrophysical sources is yet to be understood~\cite{Nagakura:2023xhc,Shalgar:2023aca,Shalgar:2024gjt}.

Understanding the quasi-steady state flavor configuration resulting from fast flavor conversion is an active field of research~\cite{Shalgar:2022rjj, Shalgar:2022lvv,Nagakura:2022xwe,Nagakura:2023mhr,Nagakura:2023xhc,Xiong:2024pue,Cornelius:2023eop,Shalgar:2019qwg,Shalgar:2024gjt,Xiong:2024tac,Martin:2021xyl,Zaizen:2023ihz,Zaizen:2022cik,Xiong:2023vcm,Richers:2021xtf,Bhattacharyya:2020jpj,Bhattacharyya:2022eed,Wu:2021uvt,Fiorillo:2024qbl,Johns:2023jjt,Nagakura:2023jfi,Johns:2024dbe}. While a robust criterion to predict the quasi-steady state flavor configuration is still missing, it is clear that the boundary conditions assumed for the QKE simulations as well as the shape of the ELN distribution could dramatically affect the final flavor configuration~\cite{Cornelius:2023eop,Shalgar:2022rjj,Shalgar:2022lvv,Shalgar:2024gjt}. On the other hand, it remains to be clarified whether the quasi-steady state flavor configuration is unique~\cite{Fiorillo:2024qbl} and how the quasi-steady state configuration could change, when relaxing the symmetry assumptions imposed on the system. This issue was investigated in the context of slow flavor conversion~\cite{Raffelt:2007yz,Raffelt:2013rqa,Mangano:2014zda}, and preliminary work focusing on fast flavor conversion shows that increasing the dimensionality of the simulation box~\cite{Richers:2021xtf,Shalgar:2019qwg} or breaking the azimuthal symmetry~\cite{Shalgar:2021oko} can lead to efficient spreading of flavor waves across neutrino angular modes and spatial regions.

In this paper, we explore the flavor conversion phenomenology in two spatial dimensions, moving beyond the assumption of spherical symmetry previously employed~\cite{Shalgar:2022rjj, Shalgar:2022lvv,Shalgar:2024gjt,Cornelius:2023eop}. Our goal is to investigate how the flavor conversion physics changes in the presence of multiple spatial dimensions. Introducing perturbations across the simulation annulus, we find that the quasi-steady state flavor configuration changes qualitatively, breaking the initial symmetry of the system. In addition, the dynamical impact of neutrino advection and collisions spreads flavor instabilities across neighboring regions in the simulation shell.

Our work is organized as follows. Section~\ref{Sec:QKE} presents the QKEs in two spatial dimensions as well as our simulation setup. Section~\ref{Sec:no_pert} focuses on our findings on the quasi-steady state flavor configuration in two spatial dimensions in the absence of perturbations. We then investigate how the flavor structure changes as a result of flavor conversion when the vacuum Hamiltonian or the collision term are perturbed in Sec.~\ref{Sec:pert}. We summarize and discuss our results in Sec.~\ref{Sec:conclusion}. Additionally, Appendix~\ref{App:adv} outlines the derivation of the advection term in two spatial dimensions, Appendix~\ref{App:conv} demonstrates the numerical convergence of our simulations, and Appendix~\ref{App:example} presents the solution of the QKEs in two spatial dimensions for a different collision term (i.e.~different ELN configuration) to validate the robustness of our findings.

\section{Neutrino quantum kinetic equations}
\label{Sec:QKE}
In this section, we introduce the neutrino QKEs. We also describe the employed simulation setup.

\subsection{Neutrino equations of motion}

\begin{figure}
\centering
\includegraphics[width=0.7\textwidth]{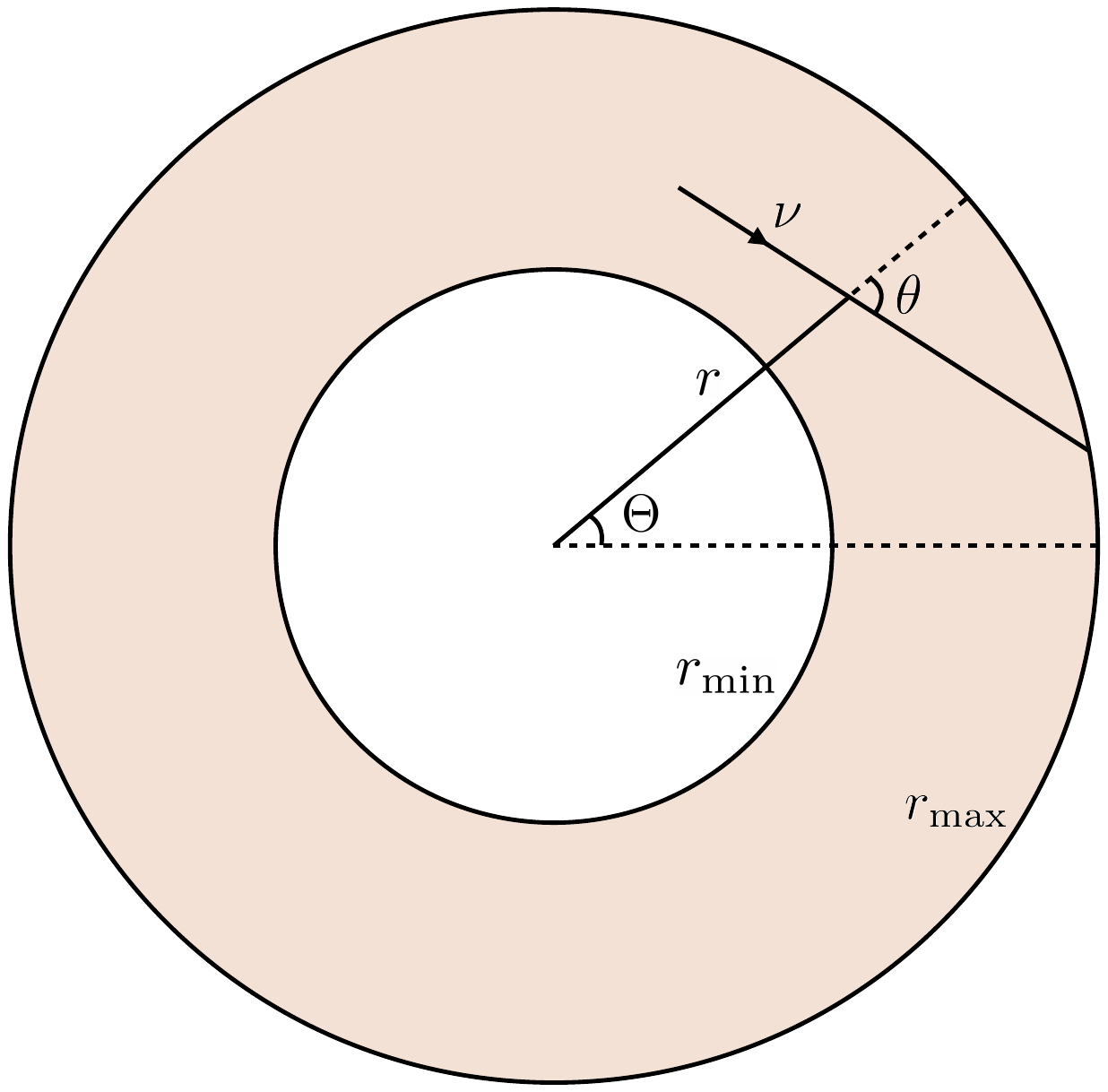}
\caption{Sketch of the shell adopted to solve the QKEs numerically. 
A point in the simulation annulus is characterized by the radius $r$, varying between the innermost (outermost) radius $r_{\min}$ ($r_{\max}$), and the polar angle $\Theta$. The neutrino direction of propagation is defined by $\theta$. 
}
\label{Fig:2D_geometry}
\end{figure}

In order to solve the neutrino QKEs in two spatial dimensions, we rely on a shell characterized by the radius $r$ and the polar angle $\Theta$, as depicted in Fig.~\ref{Fig:2D_geometry}. For each point in the simulation annulus, determined by $(r, \Theta)$, the neutrino direction of propagation is given by the angle $\theta$. Hence, the neutrino field depends on the coordinates $(r, \Theta, \theta)$, with the radius varying between $r_{\min} = 15$~km and $r_{\max}=30$~km. As for the angles, we consider $\theta \in [0,2\pi)$ and $\Theta \in [0,2\pi)$, with periodic boundaries in $\Theta$ (which means that neutrinos move across $\Theta = 0$ and $2\pi$). This choice of the simulation setup is motivated by the fact that, allowing for $\Theta \in [0, 2\pi)$, (anti)neutrinos can freely move around the annulus and we investigate an intermediate configuration with respect to a 3D one; this approach allows for a breaking of the initially spherically symmetric configuration without introducing the dependence on the azimuthal angles. 

For simplicity, we focus on two neutrino flavors $\nu_e$ and $\nu_x$, with $x=\mu$ and $\tau$. We consider $2\times 2$ density matrices, whose diagonal terms $\rho_{ii}$ (with $i= e, x$) represent the neutrino occupation numbers and are normalized such that $\rho_{ee}(r_{\min}, \Theta, \theta) = \rho_{xx}(r_{\min}, \Theta, \theta) = \bar \rho_{xx}(r_{\min}, \Theta, \theta) = 1$ and $\bar \rho_{ee}(r_{\min}, \Theta, \theta) = 1.2$. The coherence between the flavors is described by the off-diagonal terms $\rho_{ij}$, with $i \neq j$. The correspondent equations of motion for neutrinos and antineutrinos, respectively, are~\cite{Sigl:1993ctk}:
\begin{align}
  i\left(\frac{\partial}{\partial t} + \Vec{v}\cdot \vec\nabla \right) \rho(r,\Theta,\theta,t) &= \left[H,\rho(r,\Theta,\theta,t)\right] + i\mathcal{C} \label{Eq:QKEs1}\ ,\\
  i\left(\frac{\partial}{\partial t} + \Vec{v}\cdot \vec\nabla \right) \bar{\rho}(r,\Theta,\theta,t) &= \left[\bar{H},\bar{\rho}(r,\Theta,\theta,t)\right] + i\bar{\mathcal{C}}\ , \label{Eq:QKEs2}
\end{align}
where we have neglected general relativistic corrections~\cite{Nagakura:2023mhr}.
The advective term on the left-hand side of the equations of motion is derived in Appendix \ref{App:adv} and is given by
\begin{equation}
  \vec v \cdot \vec \nabla \rho(r,\Theta,\theta) = \cos\theta \frac{\partial \rho}{\partial r} + \frac{\sin\theta}{r}\frac{\partial\rho}{\partial\Theta} - \frac{\sin\theta}{r}\frac{\partial\rho}{\partial\theta}\ .
  \label{Eq:adv}
\end{equation}

The flavor conversion physics is taken into account through the commutator of the Hamiltonian and the density matrix on the right-hand side of the equations of motion. The Hamiltonian consists of the vacuum, matter, and self-interaction terms. For simplicity, we neglect the matter term and use a smaller effective vacuum mixing angle instead~\footnote{Note that a small effective mixing angle is often used in the literature to take into account the effect of matter suppression~\cite{Hannestad:2006nj}. However, in the context of a boundary problem like ours, future work should aim to assess whether the matter background can efficiently suppress flavor conversion~\cite{Esteban-Pretel:2008ovd,Dasgupta:2015iia,Abbar:2015fwa,Sigl:2021tmj}.}. Hence,   
\begin{equation}
    H=H_{\rm{vac}}+H_{\nu\nu}\ \mathrm{and}\ \bar{H}=- {H}_{\rm{vac}}+H_{\nu\nu}\ . 
\end{equation}
The vacuum term of the Hamiltonian is defined as 
\begin{equation}
    H_{\rm{vac}} = \frac{\omega}{2} \begin{pmatrix}
    -\cos 2\vartheta_{\rm{V}} & \sin 2\vartheta_{\rm{V}} \\
    \sin 2\vartheta_{\rm{V}} & \cos 2\vartheta_{\rm{V}}
    \end{pmatrix}\ ,
\end{equation}
where $\vartheta_{\rm{V}}= 10^{-3}$ is the effective vacuum mixing angle, $\omega = \Delta m^2/2E$ is the vacuum frequency ($\Delta m^2= 2.5\times 10^{-3}~\rm{eV}^2$ is the mass difference squared and $E=20$~MeV is the neutrino energy; note that we consider mono-energetic neutrinos for the sake of simplicity).
The self-interaction Hamiltonian is
\begin{equation}
    H_{\nu\nu} = ~\mu_0 \int_{0}^{2\pi} \left[\rho(r,\Theta,\theta^\prime) - \bar{\rho}(r,\Theta,\theta^\prime) \right] \times \left(1-\cos\theta\cos\theta^\prime - \sin\theta\sin\theta^\prime\right) d\theta^\prime\ ,
\label{Eq:H_nunu}
\end{equation}
where $\mu_0 = 10^4~\rm{km}^{-1}$ is the strength of the neutrino-neutrino interaction at $r_{\rm min}$ (the neutrino self-interaction strength then  decreases with radius). We note that, since we solve the neutrino QKEs in 2D, $H_{\nu\nu}$ carries an additional dependence on $\sin\theta\sin\theta^\prime$, which is not present in 1D and contributes to affect the flavor evolution.

We use a simplified model for non-forward collisions of neutrinos off matter, taking into account emission, absorption, and direction-changing scatterings (we refer the interested reader to, e.g., Refs.~\cite{Mezzacappa:2020oyq,Sigl:1993ctk,1990Ap&SS.165...65R} for overviews on the modeling of the collision term). It is given by $\mathcal{C}(r,\theta) = \mathcal{C}_{\rm{emission}} + \mathcal{C}_{\rm{absorb}} + \mathcal{C}_{\rm{dir-ch}}$, where~\cite{Shalgar:2022lvv}:
\begin{eqnarray}
    \mathcal{C}_{\text{emission}}^{i}&=&\frac{1}{\lambda_{\text {emission}}^{i}(r)}\ ,\\
    \mathcal{C}_{\text{absorb}}^{i} &=&-\frac{1}{\lambda_{\text{absorb}}^{i}(r)} \rho_{i i}(\theta)\ ,\\
    \mathcal{C}_{\text{dir-ch}}^{i} &=&-\frac{2\pi}{\lambda_{\text{dir-ch}}^{i}(r)} \rho_{i i}(\theta) + \int_{0}^{2\pi} \frac{1}{\lambda_{\text{dir-ch}}^{i}(r)} \rho_{i i}\left(\theta^{\prime}\right) d\theta^{\prime}\ .
\label{Eq:col_term}
\end{eqnarray}
Here, $\lambda^i(r)$ are the flavor-dependent mean free paths. It should be noted that the direction changing term $\mathcal{C}_{\text{dir-ch}}^{i}$, by definition, conserves the number of particles. However, this is not the case for the emission and absorption terms. 
For the modeling of the collision term, we follow Case C of Ref.~\cite{Shalgar:2022lvv}, with the mean free paths for each flavor being summarized in Table~\ref{Tab:mfp}. With this choice of the collision term, we have isotropic angular distributions for all flavors at $r_{\rm min}$, then neutrinos gradually decouple from matter as $r$ increases in a flavor-dependent fashion and ELN crossings arise (see Sec.~\ref{Sec:no_pert}).

\begin{table}[t]
    \caption{Mean free paths of emission, absorption, and direction-changing interactions for each flavor as a function of radius, with $\xi(r) = \exp(15-r)$. The modeling of the collision term follows Case C of Ref.~\cite{Shalgar:2022lvv}.}
 \centering
 \begin{tabular}{|l|l|l|l|}
 \hline & $\nu_e$ & $\bar{\nu}_{{e}}$& $\nu_x,~\bar{\nu}_x$ \\
 \hline \hline $\lambda_{\text{emission}_{\phantom{l}}}^{i}$[km] & $1 / [50 \xi(r)]$ & $1 / [30 \xi(r)]$ & $1 / [10 \xi(r)]$ \\
 \hline $\lambda_{\text {absorb }_{\phantom{l}}}^{i}$[km] & $1 / [50 \xi(r)]$ & $1 / [25 \xi(r)]$ & $1 / [10 \xi(r)]$ \\
 \hline $\lambda_{\text {dir-ch }_{\phantom{l}}}^{i}$[km] & $1 / [50 \xi(r)]$ & $1 / [25 \xi(r)]$ & $1 / [12.5 \xi(r)]$ \\
 \hline
 \end{tabular} 
 \label{Tab:mfp}
\end{table}

\subsection{Simulation setup}
The QKEs are solved relying on an adaptive multi-step Adams-Bashforth-Moulton method with adaptive step size for the time variable; the spatial and momentum derivatives are instead computed using the central difference method.
We use a uniform grid of $75$ bins for radius, angle $\Theta$, and momentum angle $\theta$. We have tested the numerical convergence of our simulations; the convergence tests for Case 1NP are reported in Appendix~\ref{App:conv}.

We aim to compute the quasi-steady state configuration that is achieved by the system in the presence of neutrino flavor conversion. In order to efficiently tackle this issue, we follow the approach presented in Refs.~\cite{Shalgar:2022lvv,Shalgar:2022rjj} and first solve Eqs.~\ref{Eq:QKEs1} and \ref{Eq:QKEs2} for $H=\bar{H}=0$ for $t=100~\mu$s to compute the ``classical steady state'' solution. 
This solution is then used as input to solve Eqs.~\ref{Eq:QKEs1} and \ref{Eq:QKEs2} when the physics of neutrino flavor conversion is taken into account~\footnote{It should be noted that the approach introduced in Refs.~\cite{Shalgar:2022lvv,Shalgar:2022rjj} is employed for numerical efficiency, but the quasi-steady state configuration is independent of the initial configuration used in the simulation. This method has the added benefit of highlighting the effect of flavor evolution on neutrino transport.}. The ``quasi-steady state'' configuration is computed solving Eqs.~\ref{Eq:QKEs1} and \ref{Eq:QKEs2} for $t=50~\mu$s. This represents the time it takes for neutrinos to cross the simulation annulus and we have tested that an approximately steady state configuration has been reached (results not shown here).

In order to compute the classical steady-state solution (as well as the quasi-steady state one), the boundary conditions of the annulus are such that neutrinos of all flavors have isotropic angular distributions at $r_{\rm min}$ and are fully decoupled from matter with a negligible backward flux at $r_{\rm max}$~\cite{Shalgar:2022lvv,Shalgar:2022rjj}. Through this approach, ELN crossings also appear within the simulation annulus when neutrinos decouple, as illustrated in the next section. 

\section{Flavor evolution in two dimensions without perturbations}
\label{Sec:no_pert}
In this section, we present the quasi-steady state neutrino flavor configuration in the absence of $\Theta$-dependent perturbations in the vacuum Hamiltonian and neutrino backgrounds. We first focus on this configuration since it allows us to recover a quasi-steady state configuration equivalent to the one obtained in a simulation shell with one spatial dimension. This is due to the symmetry across $\Theta$, which is preserved by flavor conversion to a good approximation.

\begin{figure}
\centerline{Case 1NP (no perturbations)}
\centering
\includegraphics[width=0.99\textwidth]{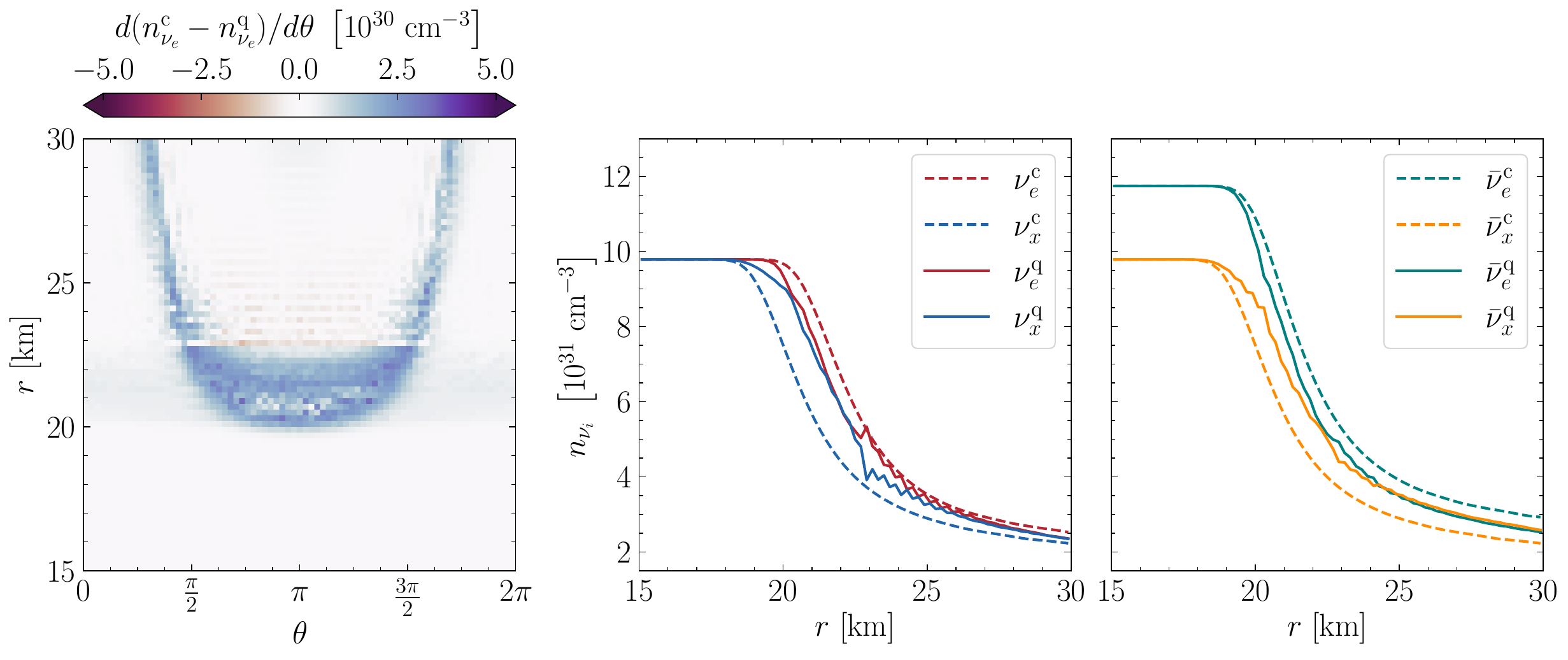}
\caption{Quasi-steady state flavor configuration obtained for the QKE solution without perturbations (Case 1NP). The simulation is evolved for $t=50~\mu$s and we show an average of the QKE solutions between $t=40~\mu$s and $t=50~\mu$s. {\it Left panel:} Contour plot of the difference of $dn_{\nu_e}/d\theta$ between the classical (c) and the quantum (q) solutions in the plane spanned by $\theta$ and $r$. The blue regions indicate a deficit of electron neutrinos due to flavor conversion.
{\it Middle and right panels:} Radial evolution of the neutrino number densities with (solid) and without (dashed) flavor conversion. The middle panel shows the number density of $\nu_e$ (red) and $\nu_x$ (blue), and the right panel shows the same for $\bar\nu_e$ (teal) and $\bar\nu_x$ (orange). The relative flavor content changes due to flavor conversion as a function of radius, but not of $\Theta$. 
}
\label{Fig:heatmap_angav_case1NP}
\end{figure}

Figure~\ref{Fig:heatmap_angav_case1NP} shows the quasi-steady state solution obtained by solving Eqs.~\ref{Eq:QKEs1} and \ref{Eq:QKEs2} with the collision term defined as in Table~\ref{Tab:mfp} (Case 1NP, no perturbations). The contour plot on the left depicts the difference in the steady state configuration of the differential electron neutrino number density ($dn_{\nu_e}/d\theta$), between the classical solution (c) and the quantum solution (q) in the plane spanned by the momentum angle $\theta$ and the radius $r$. 
We define the neutrino number density as
\begin{equation}
    n_{\nu_{i}}(r,\Theta) = \frac{\mu_0}{\sqrt{2}G_F} \int_{0}^{2\pi} \rho_{ii}(r,\Theta,\theta) d\theta \ ,
\end{equation}
with $G_F$ being the Fermi constant. Note that we do not consider the number density as the number of particles per unit area, as expected within a 2D geometry, instead choose to use the effective number of particles per unit volume at the location $(r,\Theta)$ of a 3D equivalent system (this is done rescaling by $\mu_0$). The plots in Fig.~\ref{Fig:heatmap_angav_case1NP} have been obtained for $\Theta=0^{\circ}$, however the QKE solution is identical for any $\Theta$. 

The colored regions of the contour plot on the left panel of Fig.~\ref{Fig:heatmap_angav_case1NP} show where neutrinos have changed flavor compared to the classical solution: blue (red) represents a deficit (surplus) of $\nu_e$ in the quantum solution with respect to the classical one.
At radii close to $r_{\rm min}$, neutrinos are trapped inside the neutrinosphere. As expected, we find no flavor conversions at such radii since the angular distributions of neutrinos and antineutrinos are both approximately isotropic. As the neutrinos move to larger radii, they start to decouple from the matter background and free-stream; this results in a forward-peaking of their angular distributions~\cite{Tamborra:2017ubu,Brandt:2010xa}. The electron antineutrinos decouple at smaller radii than the electron neutrinos since they have a larger mean free path (see Table~\ref{Tab:mfp}). Consequently, an ELN crossing develops which triggers flavor instabilities. This is visible through the blue band arising at $\simeq 20$~km and spreading towards larger radii in the forward direction (at $\theta=0$ and $\theta=2\pi$).

The QKE solution is independent of $\Theta$, hence Case 1NP is the analog of Case C presented in Ref.~\cite{Shalgar:2022lvv} (cf.~their Fig.~7) because ${\partial\rho}/{\partial\Theta} = 0$ in Eq.~\ref{Eq:adv}; however, note that $H_{\nu\nu}$ (Eq.~\ref{Eq:H_nunu}) acquires an additional dependence on $\sin\theta \sin\theta'$ which is not present the correspondent 1D solution. 
It can be seen from the contour plot that the quasi-steady state configuration has a left-right asymmetry around $\theta=\pi$. 
This is due to the fact that the self-interaction Hamiltonian includes a numerical integral over $\sin\theta^{\prime}$ whose floating-point error provides a seed for the spontaneous breaking of the symmetry around $\theta=\pi$.

The middle and right panels of Fig.~\ref{Fig:heatmap_angav_case1NP} show the radial profiles of the neutrino number densities for $e$ and $x$ neutrinos with and without flavor conversion, and the same for antineutrinos, for $\Theta=0^{\circ}$. 
As neutrinos decouple from matter and ELN crossings form, flavor conversion is triggered. Because of flavor conversion, the electron (anti)neutrino distribution approaches the one of the $x$ flavor. However, flavor equipartition is not achieved for neutrinos nor antineutrinos for this ELN configuration.

To better understand how neutrino flavor conversion evolves with time as a function of radius and momentum angle $\theta$, we rely on the Fourier spectrum in the momentum angle ($k_{\theta}$) for the diagonal element of the density matrix ($\rho_{ii}$): 
\begin{equation}
    \hat{F}_i(t,k_{\theta},r)=\frac{1}{2\pi}\int_{0}^{2\pi} \rho_{ii}(t,r,\Theta,\theta)e^{-ik_{\theta}\theta}d\theta  \ ,
\end{equation}
which is independent of $\Theta$. The relative contribution of the Fourier modes gives an indication of how the angular scales of the solution develop, allowing for a comparison across different radii.

\begin{figure}
\centerline{Case 1NP (no perturbations)}
\vspace{0.5cm}
\centering
\includegraphics[width=0.99\textwidth]{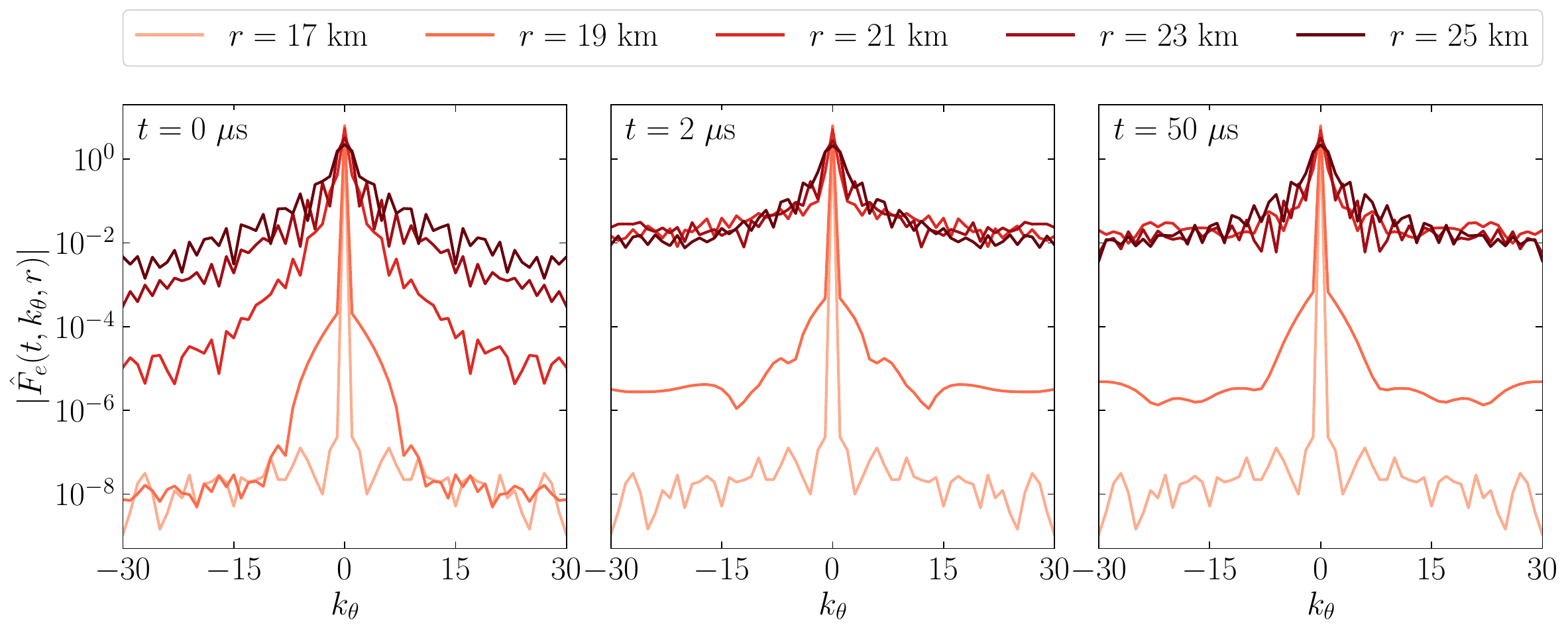}
\caption{Fourier spectra for the $\nu_e$ quasi-steady state flavor configuration obtained for the standard configuration without perturbations (Case 1NP). The Fourier spectrum for different simulation times ($t=0$, $2$, and $50~\mu$s) is shown from left to right, respectively. 
For each time, the Fourier spectrum is plotted for five selected radii ($r= 17$, $19$, $21$, $23$, and $25$~km). 
For $r \gtrsim 20$~km, where neutrinos change flavor (see Fig.~\ref{Fig:heatmap_angav_case1NP}), the contribution of the higher order Fourier modes increases for $t = 2$ and $50~\mu$s. Below $20$~km, lower order modes dominate. 
}
\label{Fig:Fourier}
\end{figure}
Figure~\ref{Fig:Fourier} shows the absolute value of the Fourier spectrum for the quasi-steady state solution of Case 1NP for $\nu_e$ for five selected representative radii ($r = 17$~km and $19$~km are below the region where flavor conversion happens, while the other three selected radii correspond to a radial region where flavor conversion takes place). The left panel shows the Fourier spectrum of the quantum solution for $t=0~\mu$s (i.e.~when there is no flavor conversion). The Fourier spectrum peaks at low order Fourier modes $k_\theta$ at small radii. For larger radii, the higher order modes are populated. This is due to the increasing forward-peaking of the angular distributions as a function of radius. 
The middle and right panels of Fig.~\ref{Fig:Fourier} show the Fourier spectra for $t=2~\mu$s and $t=50~\mu$s, respectively. At small radii, the Fourier spectrum is dominated by lower order modes $k_\theta$. As $r$ increases, higher order Fourier modes contribute to the Fourier spectrum, since neutrinos experience flavor conversion, with flavor waves cascading down to smaller angular scales. Comparing these two panels, one can see that the Fourier spectra at the selected radii are qualitatively comparable.

\section{Flavor evolution in two dimensions with perturbations}
\label{Sec:pert}
The neutrino and matter backgrounds are not homogeneous in the core of a compact astrophysical source~\cite{Tamborra:2014aua,Tamborra:2014hga,Walk:2019miz,Shibagaki:2020ksk,Takiwaki:2017tpe,Nagakura:2021lma}. In this section, we investigate the impact of such fluctuations on the flavor conversion physics solving the QKEs in the presence of $\Theta$-dependent perturbations in the vacuum term of the Hamiltonian (Case 1HP; see Sec.~\ref{Subsec:HP}) and in the neutrino background (Case 1CP; see Sec.~\ref{Subsec:CP}). 
Table~\ref{Tab:cases} shows an overview of the different configurations explored in this work.

\begin{table}
\caption{Summary of the type of the $\Theta$-dependent perturbations.}
\centering
\begin{tabular}{|l||l|}
\hline
Case 1NP & No Perturbations \\ \hline
Case 1HP & $H_{\mathrm{vac}}$ Perturbations  \\ \hline
Case 1CP & $\mathcal{C}_{\mathrm{emission}}$ Perturbations \\ \hline
\end{tabular}
\label{Tab:cases}
\end{table}

\subsection{Flavor evolution in two dimensions with perturbations in the Hamiltonian}
\label{Subsec:HP}
\begin{figure}
\centerline{Case 1HP (perturbations in vacuum Hamiltonian)}
\centering
\includegraphics[width=0.99\textwidth]{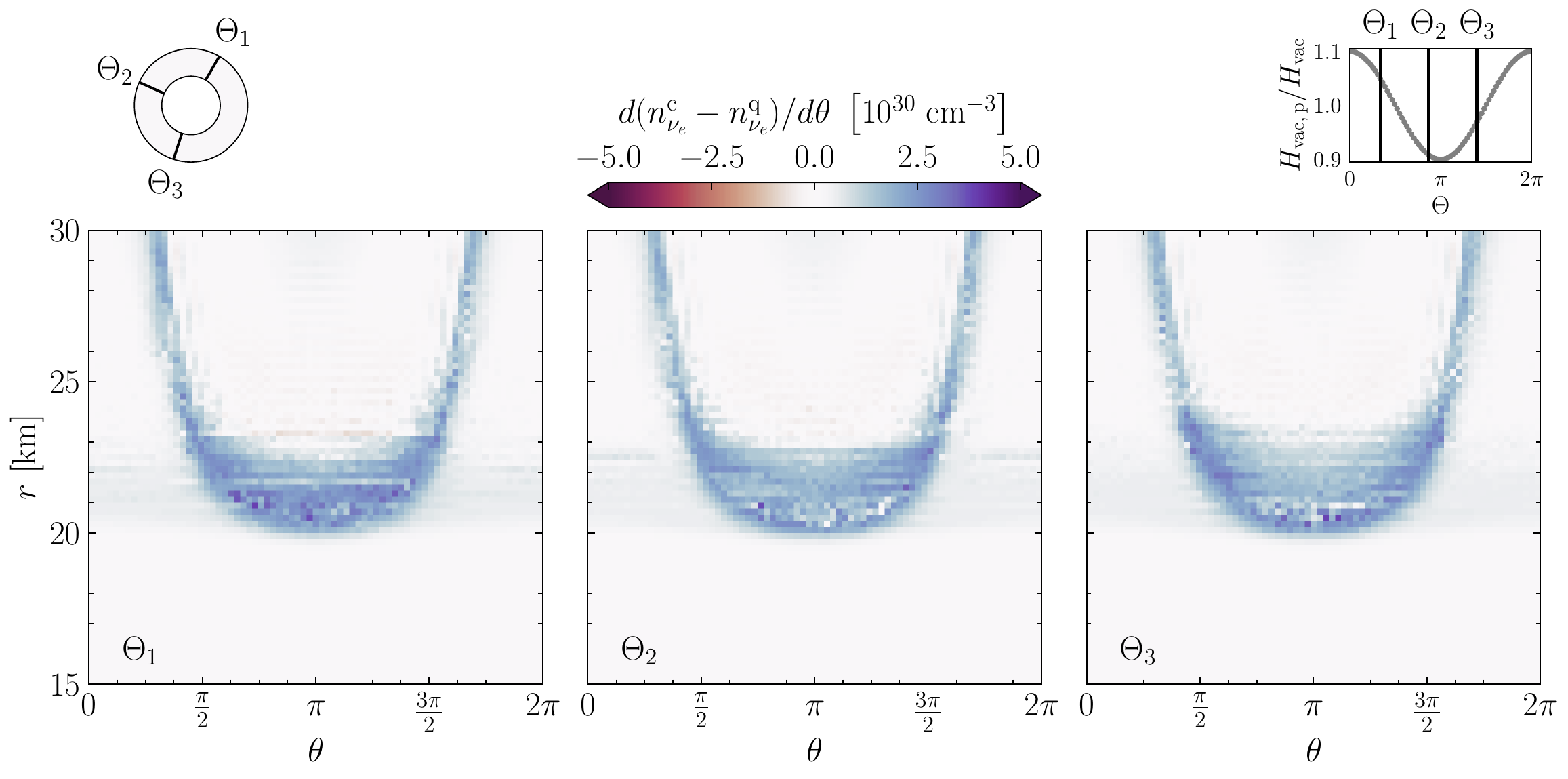}
\includegraphics[width=0.99\textwidth]{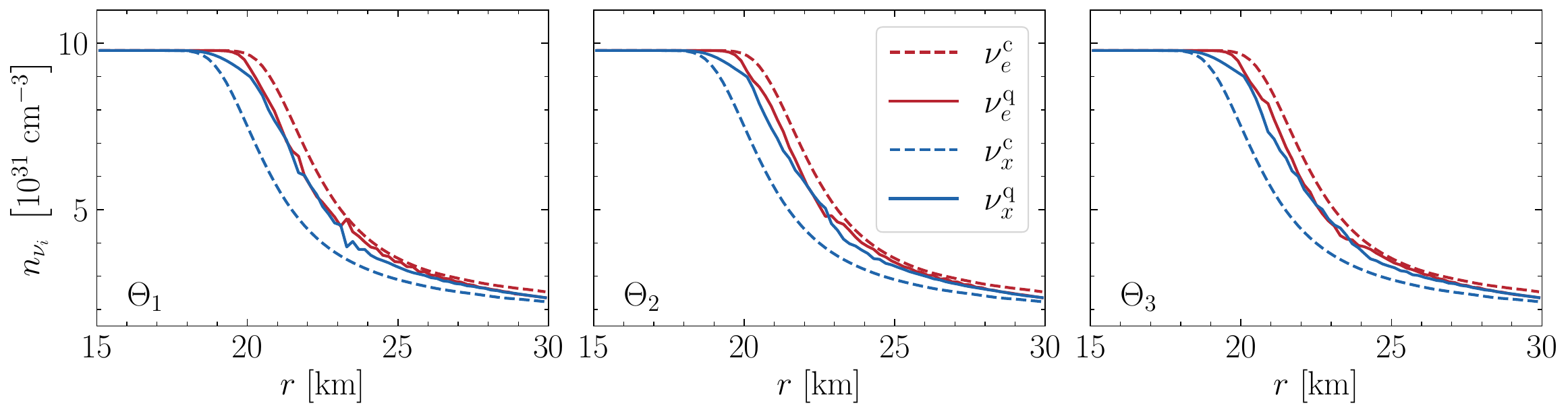}
\caption{Quasi-steady state flavor configuration obtained in the presence of perturbations in the vacuum Hamiltonian (Case 1HP, cf.~Eq.~\ref{eq:pH}). The simulation is evolved for $t=50~\mu$s; we show an average of the QKE solutions between $t=40~\mu$s and $t=50~\mu$s. {\it Top panels:} Contour plots of the difference in $dn_{\nu_e}/d\theta$ without flavor conversion and with flavor conversion in the plane spanned by the momentum angle $\theta$ and radius. Each panel represents the quasi-steady state configuration for a selected angle $\Theta$, as indicated in the inset in the upper left corner of the figure ($\Theta_1=60^{\circ}$, $\Theta_2=156^{\circ}$, and $\Theta_3=252^{\circ}$). The $\Theta$-dependence of the perturbations in the vacuum Hamiltonian is illustrated in the inset in the upper right corner. The Hamiltonian perturbations are responsible for the small scale structure differences across $\Theta$, while the overall flavor conversion trend is comparable for the different $\Theta$'s. 
{\it Bottom panels:} Radial profiles of the neutrino number densities of $\nu_e$ (red) and $\nu_x$ (blue) with (solid) and without flavor conversions (dashed) for the three selected $\Theta$ directions. While the classical steady state configurations are independent of $\Theta$, $\Theta$-dependent structures develop because of flavor conversion.
}
\label{Fig:heatmap_angav_tv_case1HP}
\end{figure}
We introduce a $\Theta$-dependent perturbation in the vacuum term of the Hamiltonian (Case 1HP):
\begin{equation}
\label{eq:pH}
    H_{\rm{vac},\,p} = H_{\rm{vac}} (1+0.1\cos{\Theta}) \ ;
\end{equation}
the perturbations $H_{\rm{vac},\,p}$ affect all flavors. The evolution of $H_{\rm{vac},\,p}$ as a function of $\Theta$ is illustrated in the inset in the top right corner of Fig.~\ref{Fig:heatmap_angav_tv_case1HP}. By perturbing the vacuum term of the Hamiltonian, we intend to explore the impact on the neutrino quasi-steady state configurations of perturbations (or fluctuations occurring in the supernova core) affecting all flavors equally. 
However, we stress that we focus on a toy model in this work and opt for an amplitude of $0.1$ to get a noticeable effect of the perturbations on the quasi-steady state configurations.

Figure~\ref{Fig:heatmap_angav_tv_case1HP} shows the quasi-steady state solution for Case 1HP. The contour plots in the top row show the difference in $dn_{\nu_e}/d\theta$ without and with flavor conversion in the plane spanned by the radius and momentum angle $\theta$, for three different $\Theta$ angles ($\Theta_1$, $\Theta_2$, and $\Theta_3$) from left to right, respectively. 
The three selected $\Theta$ angles are marked in the inset on the top left corner of the figure to guide the eye. 
The bottom panels of Fig.~\ref{Fig:heatmap_angav_tv_case1HP} represent the corresponding radial evolution of the neutrino number densities for $\nu_e$ and $\nu_x$ for the classical solution and quantum solutions.

Comparing Fig.~\ref{Fig:heatmap_angav_tv_case1HP} (Case 1HP) with Fig.~\ref{Fig:heatmap_angav_case1NP} (Case 1NP), one can see that the perturbations introduced in the Hamiltonian break the symmetry across the simulation annulus, leading to the development of small scale structures and different radial evolution of $n_{\nu_i}(r,\Theta)$. Because of neutrino advection and collision, the impact of perturbations is not limited to the specific $\Theta$ where they are introduced, rather it spreads across neighbor cells, in agreement with the findings of Refs.~\cite{Shalgar:2019qwg,Padilla-Gay:2020uxa,Xiong:2024tac,Nagakura:2023xhc}. 

The impact of the perturbations introduced in Eq.~\ref{eq:pH} is investigated in Appendix~\ref{App:example} for another set of neutrino angular distributions (cf.~Fig.~\ref{Fig:heatmap_angav_case2NP} vs.~\ref{Fig:heatmap_angav_tv_case2HP}). With respect to the example presented in this section, we find a qualitatively similar trend for what concerns the changes induced by flavor conversion across $\Theta$.

\subsection{Flavor evolution in two dimensions with perturbations in the collision term}
\label{Subsec:CP}
\begin{figure}
\centerline{Case 1CP (perturbations in collision term)}
\centering
\includegraphics[width=0.999\textwidth]{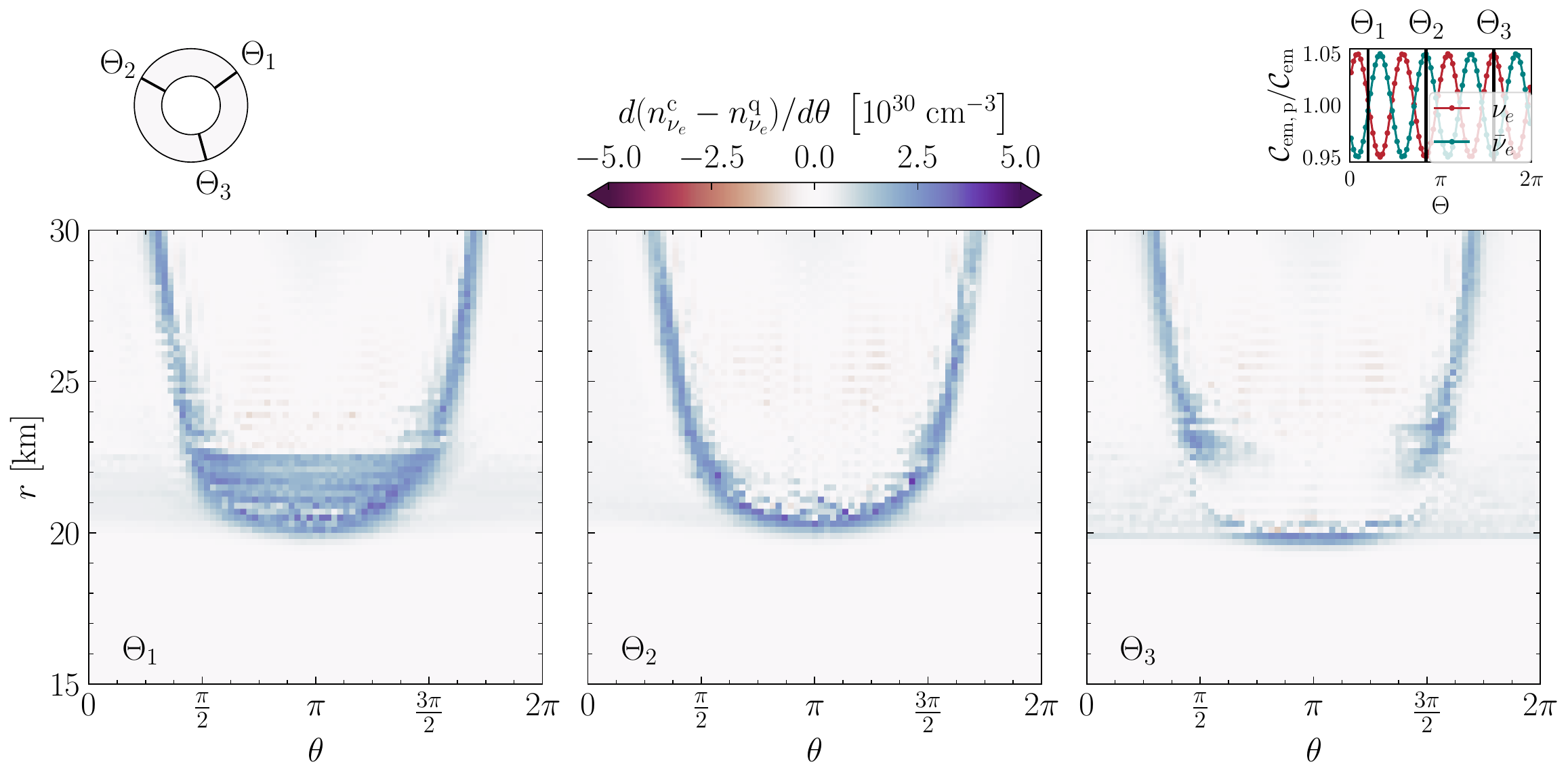}
\includegraphics[width=0.999\textwidth]{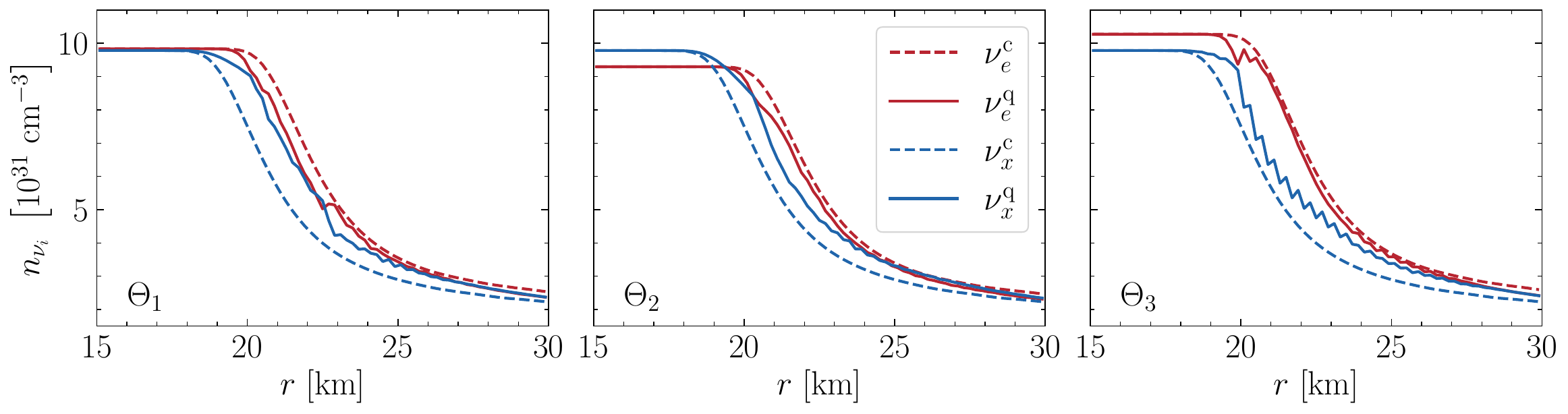}
\caption{Same as Fig.~\ref{Fig:heatmap_angav_tv_case1HP}, but for the configuration obtained introducing perturbations in the collision term (Case 1CP, Eqs.~\ref{eq:colle} and \ref{eq:collebar}). The simulation is evolved for $t=50~\mu$s; we show an average of the QKE solutions between $t=40~\mu$s and $t=50~\mu$s. 
From left to right, each panel refers to the angle $\Theta$ marked in the inset in the upper left corner ($\Theta_1 = 36^{\circ}$, $\Theta_2=151.2^{\circ}$, and $\Theta_3=285.6^{\circ}$). The quasi-steady state configuration varies according to $\Theta$ because of the perturbations that affect the flavor conversion physics. The perturbations in the collision term are modeled as indicated in the inset in the top right corner of the figure for $\nu_e$ (red) and $\bar\nu_e$ (teal). 
For $\Theta_1$, the quasi-steady state configuration resembles the one found for Case 1NP with a larger region of flavor conversion between $20$ and $23$~km. For $\Theta_2$, a smaller region of flavor conversion appears between $20$ and $21$~km, and for $\Theta_3$ the variation of $\nu_e$ due to flavor conversion is negligible for $r \simeq 21$~km. These differences are also visible in the radial evolution of the neutrino number densities plotted in the bottom panels. 
}
\label{Fig:heatmap_angav_ThC_case1CP}
\end{figure}

While Case 1HP focuses on flavor-independent perturbations, in this section we consider perturbations in the emission term of electron neutrinos and antineutrinos (with no perturbations in the vacuum term of the Hamiltonian), i.e.~we perturb the system in a flavor-dependent fashion. For electron neutrinos: 
\begin{equation}
\label{eq:colle}
    \mathcal{C}_{\rm{emission},\,p}^e = \mathcal{C}_{\rm{emission}}^e [1+0.05\cos({4\Theta-\pi/3})]\ ;
\end{equation}
while for electron antineutrinos, we consider
\begin{equation}
\label{eq:collebar}
    \bar{\mathcal{C}}_{\rm{emission},\,p}^{e} = \bar{\mathcal{C}}_{\rm{emission}}^{e} \left[1+0.05\cos({4\Theta+2\pi/3})\right] \ .
\end{equation}
The functional form of the perturbations for $\nu_e$ and $\bar\nu_e$ is illustrated in the inset in the top right corner of Fig.~\ref{Fig:heatmap_angav_ThC_case1CP}.
This type of perturbation is implemented to mimic fluctuations in the matter background which would affect the emission of electron-type neutrinos~\cite{Tamborra:2014aua}. However, in our toy-model, we choose an amplitude of $0.05$ of the harmonic variation of the emission term to be able to appreciate a breaking of the symmetry in $\Theta$. In addition, the perturbation functional form is anti-symmetric around $\Theta=\pi$ to further distinguish this type of perturbation from Case 1HP.

\begin{figure}
\centerline{Case 1CP (perturbations in collision term; classical solution)}
\vspace{0.5cm}
\centering
\includegraphics[width=0.999\textwidth]{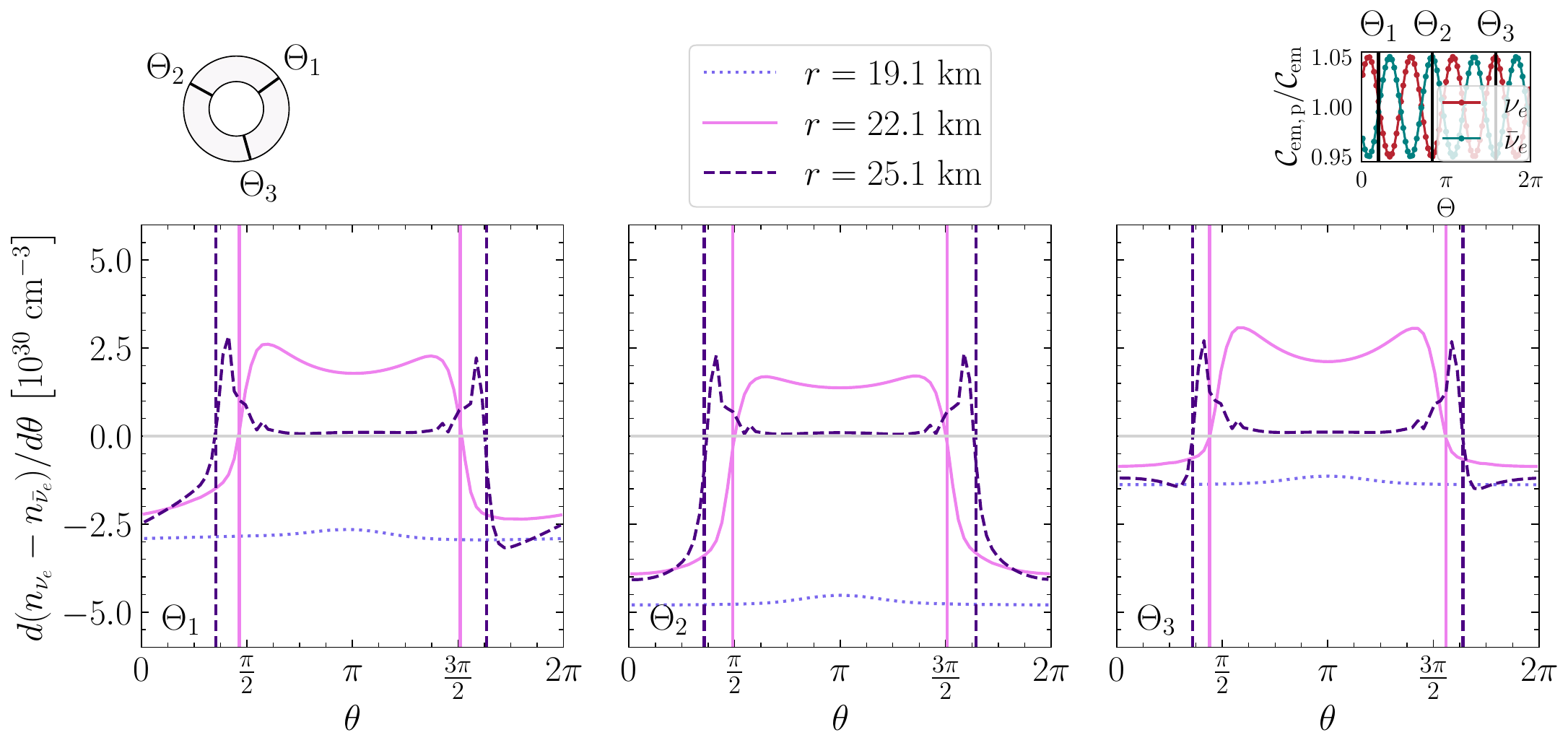}
\caption{Angular distributions of the ELN number density for the classical steady state of Case 1CP as a function of $\theta$ 
for three different radii: $r=19.1$ km (dotted, violet), $r=22.1$ km (solid, pink), and $r=25.1$ km (dashed, indigo). Each panel refers to a specific angle $\Theta$, as indicated in the inset in the top left corner. The loci of ELN crossings are marked by a gray horizontal line to guide the eye. The inset in the top right corner of the figure illustrates the modeling of the perturbation as a function of $\Theta$. Because of the different ELN configuration as a function of $\Theta$ induced by the perturbation of the collision term, the quasi-steady state neutrino number densities plotted in Fig.~\ref{Fig:heatmap_angav_ThC_case1CP} exhibit a strong dependence on $\Theta$. 
}
\label{Fig:ang_dist_ThC_case1CP}
\end{figure}
Figure~\ref{Fig:heatmap_angav_ThC_case1CP} shows the quasi-steady state configuration for Case 1CP for three selected angles $\Theta_1$, $\Theta_2$, and $\Theta_3$ 
(cf.~the inset in the top left corner of the figure), respectively chosen to represent a part of the perturbation spectrum where the $\nu_e$ and $\bar\nu_e$ emissions are comparable, a part where the $\nu_e$ emission dominates, and where the $\bar\nu_e$ emission dominates. The top panels display the difference in the (quasi-)steady state configurations for the classical and quantum solutions. 
For $\Theta_1$, a configuration similar to Case 1NP is found, with a broad region of flavor conversion between $20$ and $23$~km. For $\Theta_2$, a less extended region of flavor conversion appears (between $20$ and $21$~km), and for $\Theta_3$ the classical and quantum solutions are comparable in the proximity of $r \simeq 21$~km (cf.~white band).
The significant differences, as a function of $\Theta$, obtained in the flavor configuration after flavor conversion imply that flavor waves have not spread enough to homogenize the neutrino distribution across the simulation annulus. Yet, one can see that flavor structures spread across neighbor bins in the simulation annulus. 
This trend is also evident from the radial profiles of the neutrino number densities in the bottom panels of Fig.~\ref{Fig:heatmap_angav_ThC_case1CP}: for $\Theta_1$, $\nu_e^q$ and $\nu_x^q$ fluctuate close to each other, whereas for $\Theta_3$ they are far from each other. For $\Theta_2$ and $\Theta_3$, the perturbation is such that the radial distribution achieved in the classical steady state is different between $\nu_e$ and $\nu_x$ at small radii.

To get a better understanding of the $\Theta$-dependence of the flavor evolution, Fig.~\ref{Fig:ang_dist_ThC_case1CP} shows the ELN angular distributions of the classical steady state 
for three selected radii. At $r=19.1$ km, the angular distributions of $\nu_e$ and $\bar\nu_e$ are isotropic and do not cross, hence no fast flavor instability is expected. 
At $r=22.1$~km and $r=25.1$ km, two ELN crossings appear. 
The difference in the location of the ELN crossings for different $\Theta$ directions are responsible for the development of flavor instabilities in different spatial regions, as shown in the top panels of Fig.~\ref{Fig:heatmap_angav_ThC_case1CP}.

Appendix~\ref{App:example} provides another example of the quasi-steady state configuration achieved in the presence of perturbations in the collision term for a different flavor configuration (cf.~Fig.~\ref{Fig:heatmap_angav_ThC_case2CP}). Similar conclusions are achieved also in this case for what concerns the impact of perturbations on the quasi-steady state configuration across $\Theta$ directions. In fact, the difference in the location and extension of flavor conversion is more pronounced in this case, with the initial $\theta$ angular distributions exhibiting no ELN crossings for selected $\Theta$'s, but the spread of flavor waves contributes to the development of flavor conversion (cf.~Fig.~\ref{Fig:ang_dist_ThC_case2CP}).

\section{Discussion and outlook}
\label{Sec:conclusion}
We have presented the solution of the QKEs in $(2+1+1)$ dimensions, including two spatial dimensions, one neutrino momentum angle (focusing on mono-energetic neutrinos), and time. The neutrino angular distributions are computed through neutral and charged current interaction terms entering the collisional kernel in the QKEs. We have tracked the neutrino flavor evolution during neutrino decoupling from matter (with isotropic angular distributions at the innermost radius, that consistently evolve becoming forward peaked as the radius increases). 
As already pointed out in Refs.~\cite{Shalgar:2019qwg,Padilla-Gay:2020uxa,Shalgar:2024gjt,Shalgar:2020wcx,Nagakura:2023xhc,Xiong:2024tac,Xiong:2022vsy,Nagakura:2023wbf,Nagakura:2022qko,Martin:2021xyl,Johns:2021qby,Kato:2021cjf}, the non-trivial interplay among advection, collisions, and flavor conversion shapes the quasi-steady state configuration achieved by neutrinos.

\begin{figure}
\centering
\includegraphics[width=0.99\textwidth]{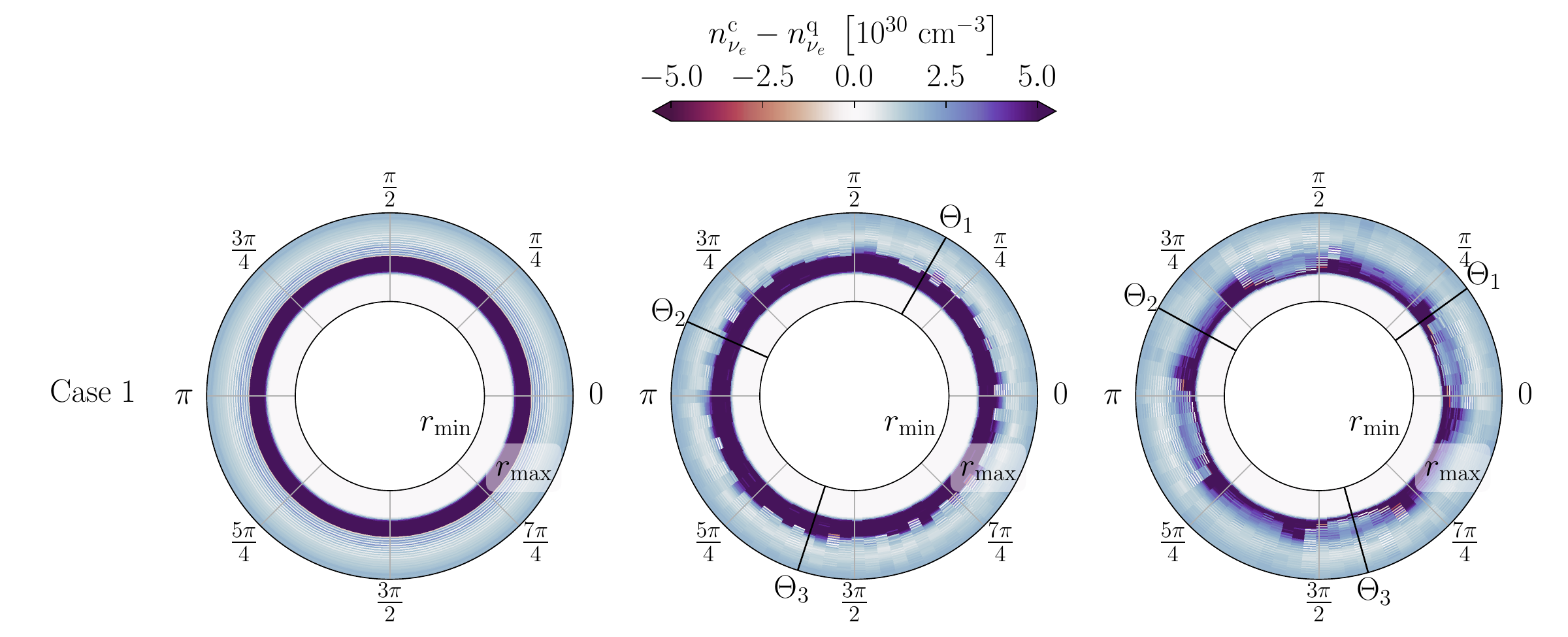}
\includegraphics[width=0.99\textwidth]{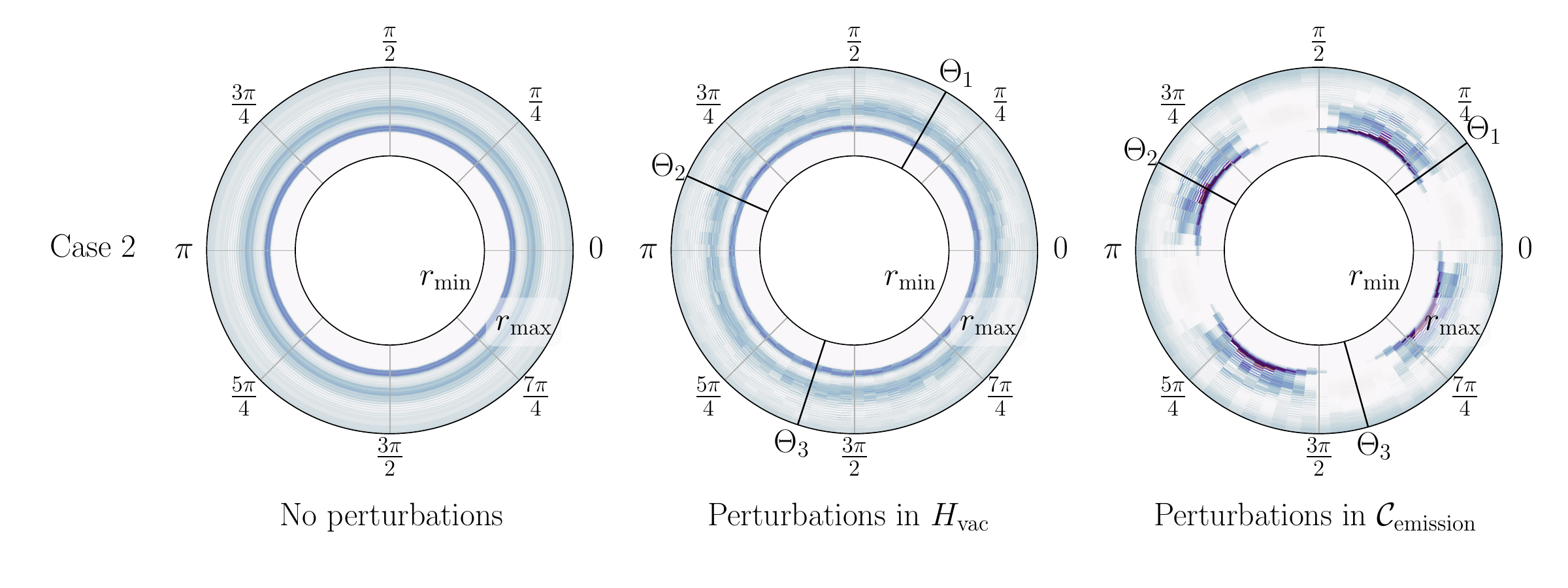}
\caption{Difference in the $\nu_e$ number density of the steady state configurations obtained without flavor conversion (c) and the quasi-steady state configurations with flavor conversion (q) for Cases NP (left, without perturbations), PH (middle, with perturbations in the vacuum Hamiltonian), and PC (right, with perturbations in the collision term). The top (bottom) panels refer to Case 1 (Case 2; cf.~Appendix~\ref{App:example}). The simulations are evolved for $t=50~\mu$s; we show an average of the QKE solutions between $t=40~\mu$s and $t=50~\mu$s. The perturbations in the Hamiltonian or in the collision term contribute to breaking the symmetry in $\Theta$, giving rise to $\Theta$-dependent quasi-steady state configurations, with flavor waves marginally spreading across neighbor $\Theta$ regions (see \href{https://sid.erda.dk/share_redirect/A8yryUX901/index.html}{animations}).
}
\label{Fig:polar_heatmap_comp}
\end{figure}
Figure~\ref{Fig:polar_heatmap_comp} summarizes our findings for the $\nu_e$ quasi-steady state configurations for Case 1 (top panels) and Case 2 (bottom panels, presented in Appendix~\ref{App:example} and obtained for a collision term leading to different $\bar\nu_e$ emission with respect to Case 1). Animations of the flavor evolution for all configurations investigated in this work are available \href{https://sid.erda.dk/share_redirect/A8yryUX901/index.html}{here}. At the innermost radii, the angular distributions for all flavors are isotropic, and flavor conversion is triggered above $\simeq 20$~km. 
In the absence of any perturbations (left panels) and despite the difference in the neutrino self-interaction Hamiltonian between the 2D and 1D simulation setups, our findings are qualitatively comparable with Cases C and B presented in Ref.~\cite{Shalgar:2022lvv} (except for a small left-right asymmetry in $\theta$ appearing in 2D). This implies that the quasi-steady state flavor configuration is independent of $\Theta$ to a good approximation.

In order to mimic fluctuations occurring in the neutrino and matter backgrounds, we introduce perturbations in the vacuum term of the Hamiltonian (affecting all flavors equally) or in the collision term (in the emission term of $\nu_e$ and $\bar\nu_e$). In both scenarios, the presence of perturbations contributes to break the symmetry in $\Theta$, giving rise to $\Theta$-dependent quasi-steady state configurations.

When all flavors are perturbed in the same way in $H_{\rm{vac}}$ (Figs.~\ref{Fig:heatmap_angav_tv_case1HP}, \ref{Fig:heatmap_angav_tv_case2HP} and middle panels of Fig.~\ref{Fig:polar_heatmap_comp}), the ELN crossings in the classical steady state configuration occur at the same momentum angle $\theta$, for all $\Theta$. Hence, we roughly obtain the same regions of flavor transformation on the contour plots as in the non-perturbed case (cf.~left panels of Fig.~\ref{Fig:polar_heatmap_comp}), except for small fluctuations.
For Case 1HP, the difference between the flavor configurations obtained for different $\Theta$'s is more pronounced than for Case 2HP.

Introducing flavor-dependent perturbations in the collision term (Figs.~\ref{Fig:heatmap_angav_ThC_case1CP} and \ref{Fig:heatmap_angav_ThC_case2CP} and right panels of Fig.~\ref{Fig:polar_heatmap_comp}), the symmetry is clearly broken in $\Theta$ because the ELN crossings occur at different $\theta$'s for each $\Theta$. Notably, because of neutrino advection, flavor waves spread through their neighbor angular regions, remaining localized in specific ranges of $\Theta$.
This trend is more pronounced in Case 2CP, where there is no crossing for some $\Theta$'s in the initial angular distribution, but advection favors the spread of the regions affected by flavor conversion (Fig.~\ref{Fig:ang_dist_ThC_case2CP}). These results suggest a larger variation of the flavor conversion physics across different $\Theta$ regions when the number of emitted neutrino flavors is not homogeneous. 

In none of the explored ELN configurations, flavor conversion leads to flavor equipartition, despite Refs.~\cite{Xiong:2024tac,Xiong:2024pue,Martin:2021xyl,Zaizen:2023ihz,Zaizen:2022cik,Xiong:2023vcm,Grohs:2022fyq,Richers:2021xtf,Richers:2022bkd,Bhattacharyya:2020jpj,Bhattacharyya:2022eed,Wu:2021uvt,Nagakura:2023xhc} conclude that this might be a generic flavor outcome in 1D on one side of the ELN crossing. In this respect, our findings are in agreement with the 1D results of Refs.~\cite{Shalgar:2022lvv,Shalgar:2022rjj}, the achievement of flavor equipartition being dependent on the choice of the simulation boundary conditions and the shape of the ELN distributions~\cite{Cornelius:2023eop}.

Earlier work attempted to explore the neutrino flavor conversion physics in multi-dimensions~\cite{Padilla-Gay:2020uxa,Shalgar:2019qwg,Richers:2021xtf}. Here, for the first time, we consistently evolve the (anti)neutrino angular distributions through the collision term, tracking neutrino decoupling from matter. Also, we do not rely on the employment of seeds in the off-diagonal terms of the density matrix to aid the growth of flavor instabilities, but evolve the QKEs including the vacuum term in the Hamiltonian. Our main findings are in general agreement with the ones reported in previous multi-D work within simplified setups, concerning the fact that advection allows for the spreading of flavor structures within small spatial regions. However, our dynamical modeling of the ELN distributions does not lead to flavor equilibration as an outcome of flavor conversion. Future work should also focus on the interplay between the increased dimensionality of the simulation shell and the inclusion of the azimuthal angle in the neutrino distribution---the inclusion of the latter has been shown to lead to spontaneous symmetry breaking~\cite{Shalgar:2021oko}. 

In summary, this work highlights the non-trivial interplay among neutrino flavor conversion, collisions, and advection in multi-dimensions, showing that flavor conversion can spread across neighbor regions of the simulation shell through neutrino advection. Our findings suggest that it is important to assess how the quasi-steady state flavor configuration changes when relaxing symmetry assumptions in the solution of the QKEs in order to work out robust criteria to forecast the flavor configuration resulting from flavor conversion, and its consequent impact on the physics of compact astrophysical sources and related multi-messenger observables.

\acknowledgments
We thank Damiano Fiorillo, Manuel Goimil Garc\'ia, Georg Raffelt, G\"unter Sigl, Meng-Ru Wu, and Zewei Xiong for their comments on the manuscript.
This project has received support from the Villum Foundation (Project No.~13164), the Danmarks Frie Forskningsfond (Project 
No.~8049-00038B), the European Union (ERC, ANET, Project No.~101087058), and the Deutsche Forschungsgemeinschaft through Sonderforschungbereich SFB 1258 ``Neutrinos and Dark Matter in Astro- and Particle Physics'' (NDM). 
Views and opinions expressed are those of the authors only and do not necessarily reflect those of the European Union or the European Research Council. Neither the European Union nor the granting authority can be held responsible for them. The Tycho supercomputer hosted at the SCIENCE HPC Center at the University of Copenhagen was used for supporting the numerical simulations presented in this work.

\appendix
\section{Neutrino advection term in two spatial dimensions}
\label{App:adv}

\begin{figure}
\centering
\includegraphics[width=0.7\textwidth]{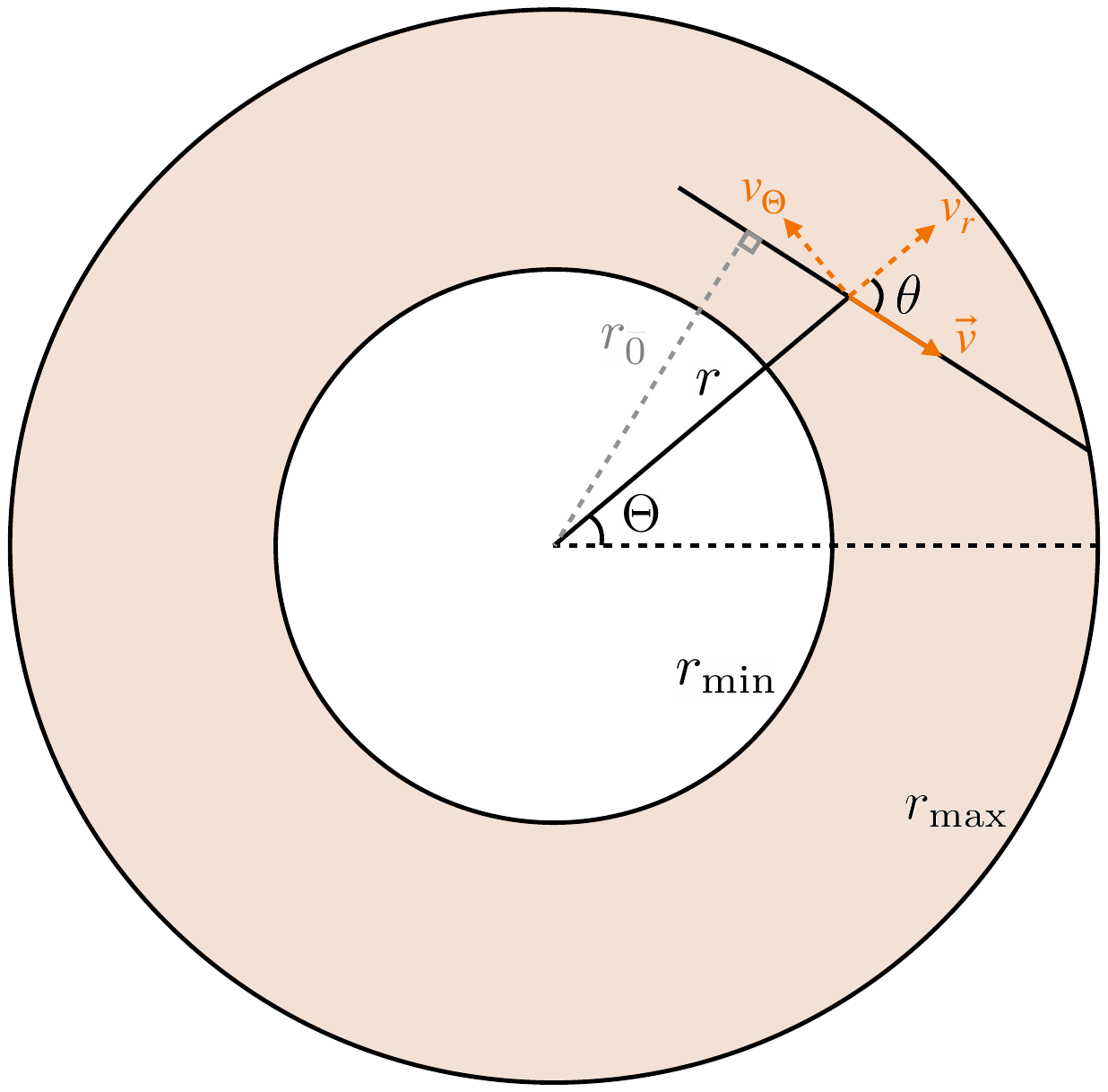}
\caption{Sketch of the simulation annulus (cf.~Fig.~\ref{Fig:2D_geometry}). A neutrino traveling with velocity $\vec v$ in the simulation annulus is described by the radius $r$, angle $\Theta$, and momentum angle $\theta$. The minimum distance between the neutrino trajectory and the center is given by $r_0$. The velocity vector can be decomposed into the radial and angular components $v_r$ and $v_{\Theta}$.
}
\label{Fig:2D_geometry_derivation}
\end{figure}
In this appendix, we derive the expression for the advection term ($\vec{v}\cdot \vec\nabla \rho(r,\Theta,\theta)$) in Eqs.~\ref{Eq:QKEs1} and \ref{Eq:QKEs2}.
The del operator in two dimensions and polar coordinates is
\begin{equation}
    \vec{\nabla} = \frac{\partial}{\partial r}\hat{r} + \frac{1}{r}\frac{\partial}{\partial \Theta}\hat{\Theta} \ ;
\end{equation}
while the velocity vector is defined as
\begin{equation}
    \vec{v} = v_r \hat{r} + v_{\Theta}\hat{\Theta} = \frac{\partial r}{\partial t} \hat{r} + r \frac{\partial \Theta}{\partial t}\hat{\Theta} \ .
\end{equation}
The velocity components can be expressed in terms of the neutrino momentum angle $\theta$ considering that $\partial r / \partial t = |\vec v| \cos\theta = \cos\theta$ and $\partial \Theta/\partial t = |\vec v| \sin\theta/r = \sin\theta/r$, since $|\vec v| \approx c = 1$ (see Fig.~\ref{Fig:2D_geometry_derivation}).
Hence 
\begin{equation}
    \vec v \cdot \vec \nabla = \cos{\theta}\frac{\partial}{\partial r} + \frac{\sin{\theta}}{r}\frac{\partial}{\partial \Theta} \ .
\end{equation}
We can use the setup illustrated in Fig.~\ref{Fig:2D_geometry_derivation} to express the neutrino momentum angle as a function of the radius. It is given by
\begin{equation}
    \cos\theta(r) = \frac{\sqrt{r^2-r_0^2}}{r} \ ,
\end{equation}
with $r_0$ being the the minimum distance to the neutrino trajectory.
Using the expressions above, the advection term in Eqs.~\ref{Eq:QKEs1} and \ref{Eq:QKEs2} becomes 
\begin{equation}
\begin{split}
    \vec v \cdot \vec \nabla \rho(r,\Theta,\theta(r)) =& \cos{\theta} \left(\frac{\partial\rho(r,\Theta,\theta(r))}{\partial r} + \pdv{\theta (r)}{r}\frac{\partial \rho(r,\Theta,\theta(r))}{\partial\theta(r)} \right) +\frac{\sin{\theta(r)}}{r} \frac{\partial\rho(r,\Theta,\theta(r))}{\partial\Theta} \ . 
    \label{Eq:adv_derivs}
\end{split}
\end{equation}
To solve Eq. \ref{Eq:adv_derivs}, we calculate the radial derivative of $\cos\theta(r)$:
\begin{equation}
    \pdv{\cos\theta(r)}{r} = \frac{r_0^2}{r^2\sqrt{r^2-r_0^2}} = \frac{r^2\sin^2{\theta(r)}}{r^2\sqrt{r^2-r^2\sin^2\theta(r)}} = \frac{\sin^2\theta(r)}{r\cos\theta(r)} \ ,
\end{equation}
such that
\begin{equation}
\label{eq:dthetadr}
    \pdv{\theta(r)}{r} = -\frac{\sin\theta(r)}{r\cos\theta(r)} \ .
\end{equation}
Inserting Eq.~\ref{eq:dthetadr} in Eq.~\ref{Eq:adv_derivs}, we obtain
\begin{equation}
\begin{split}
    \vec v \cdot \vec \nabla \rho(r,\Theta,\theta)
    = \cos\theta \pdv{\rho(r,\Theta,\theta)}{r} + \frac{\sin\theta}{r}\pdv{\rho(r,\Theta,\theta)}{\Theta} - \frac{\sin\theta}{r}\pdv{\rho(r,\Theta,\theta)}{\theta}\ ,
    \label{eq:2D_ad}
\end{split}
\end{equation}
which is the advection term used in Eqs.~\ref{Eq:QKEs1} and \ref{Eq:QKEs2}.

\section{Numerical convergence}
\label{App:conv}
\begin{figure}
\centerline{Case 1NP (no perturbations)}
\centering
\centering
\includegraphics[width=0.99\textwidth]{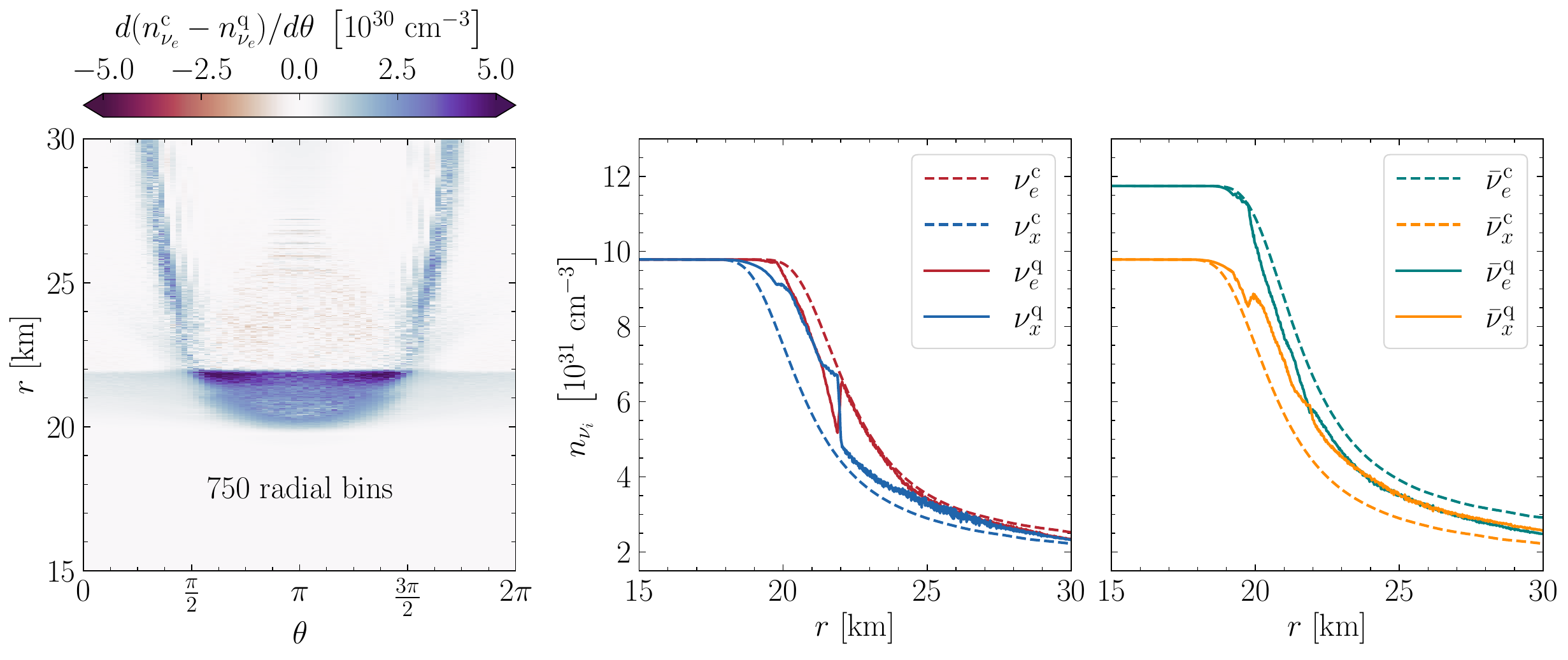}
\includegraphics[width=0.99\textwidth]{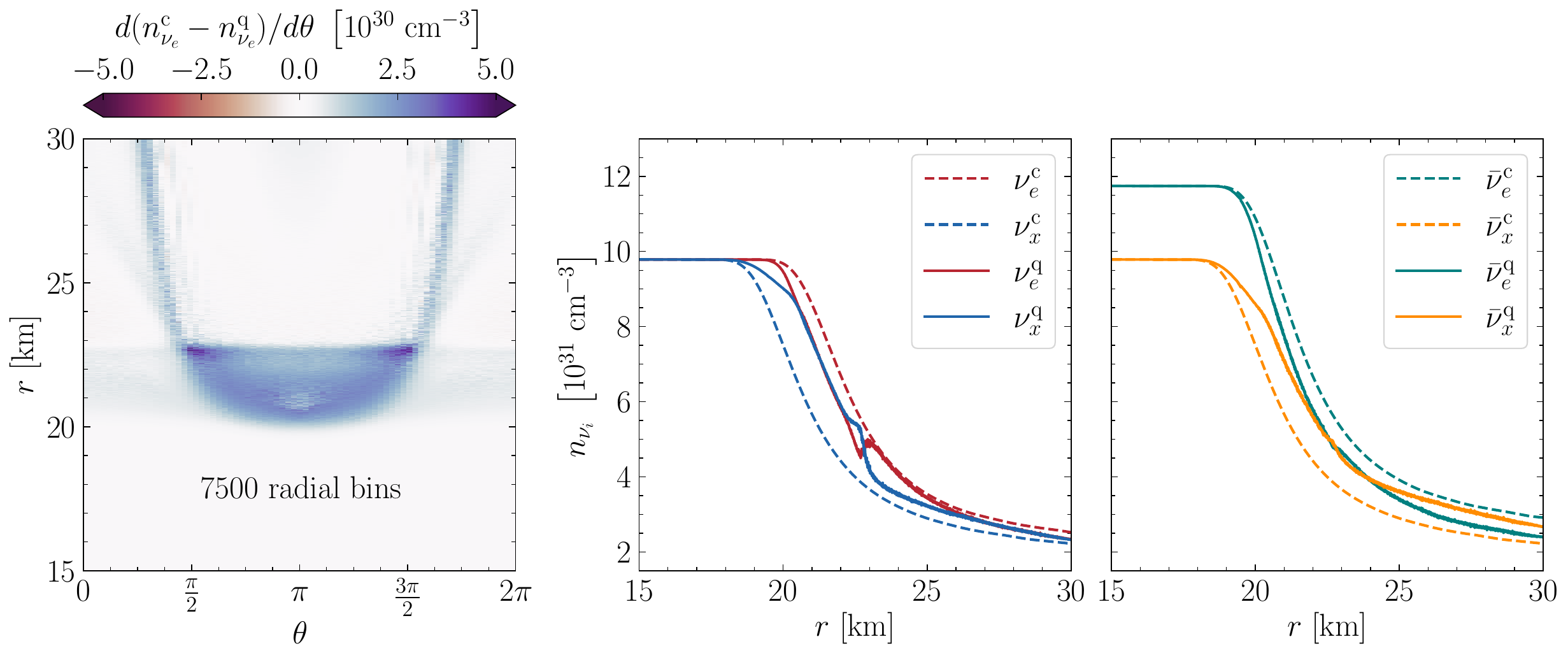}
\caption{Quasi-steady state flavor configuration obtained for the QKE solution without perturbations (Case 1NP) for 750 radial bins (top) and 7500 radial bins (bottom). The simulations are evolved for $t=50~\mu$s and plots are obtained for an average between the solution at $t=40~\mu$s and the solution at $t=50~\mu$s. {\it Left panel:} Contour plot of the difference of $dn_{\nu_e}/d\theta$ between the classical (c) and the quantum (q) solutions in the plane spanned by $\theta$ and $r$. The blue regions indicate a deficit of electron neutrinos due to flavor conversion.
{\it Middle and right panels:} Radial evolution of the neutrino number densities with (solid) and without (dashed) flavor conversion. The middle panel shows the number density of $\nu_e$ (red) and $\nu_x$ (blue), and the right panel shows the same for $\bar\nu_e$ (teal) and $\bar\nu_x$ (orange). The relative flavor content changes due to flavor conversion as a function of radius, but not of $\Theta$. 
}
\label{Fig:heatmap_angav_case1NP_convergence}
\end{figure}

In order to demonstrate numerical convergence, we present the results obtained using various radial resolutions for Case 1NP (without any perturbations).  Figure~\ref{Fig:heatmap_angav_case1NP_convergence} shows the quasi-steady state configuration obtained for $750$ radial bins (top) and $7500$ radial bins (bottom) for $\Theta=0^{\circ}$. Comparing Fig.~\ref{Fig:heatmap_angav_case1NP_convergence} to Fig.~\ref{Fig:heatmap_angav_case1NP} (the latter being carried out with $75$ radial bins), we can see a comparable trend and similar development of flavor instabilities above $20$~km for all three simulations (left panels). The radial profiles of the neutrino number densities (right panels) show that flavor equipartition is not achieved in any of these runs, albeit we see a clearer separation and subsequent crossing between the densities of $\nu_e$ and $\nu_x$ in Fig.~\ref{Fig:heatmap_angav_case1NP_convergence}. The results shown in this section are representative of other cases explored in the paper which exhibit similar trends.

As is the case for the simulation with $75$ radial bins, the contour plot with $750$~radial bins displays small scale features due to the floating-point error (cf.~red spots developing in the backwards direction above $22$~km, top panel of Fig.~\ref{Fig:heatmap_angav_case1NP_convergence}). These structures are not present for the simulation with $7500$ radial bins, but they do not have any impact on the overall flavor evolution. 

We conclude that the simulation with $75$ radial bins reliably allows to track the overall flavor phenomenology and the under-resolved small scale features do not have any impact on the global features that we aim to investigate. 
We stress that, as illustrated in Refs.~\cite{Shalgar:2022rjj,Shalgar:2022lvv}, the characteristic length scale of our system is not defined by $1/\mu_0$. In fact, in our simulation setup neutrinos are not emitted from a single surface, rather the decoupling region is extended as it is determined by the collision terms. We also include a non-zero vacuum frequency in the solution of the neutrino equations of motion, as well as neutrino advection and collisions. 

\section{Flavor evolution in two dimensions with perturbations: another electron neutrino lepton number angular configuration}
\label{App:example}
In order to assess the robustness of our findings for what concerns the role of the perturbations on the quasi-steady state flavor configurations, in this appendix, we investigate the flavor conversion physics for another set of neutrino angular distributions.
To this purpose, we consider the mean free paths defined as in Table~\ref{Tab:mfp_2}, following Case B of Ref.~\cite{Shalgar:2022lvv}. Table~\ref{Tab:cases_app} shows an overview of the cases explored in this Appendix.
\begin{table}[t]
    \caption{Mean free paths of emission, absorption, and direction-changing interactions for each flavor as a function of radius. This model for the collision term follows Case B of Ref.~\cite{Shalgar:2022lvv}.}
 \centering
 \begin{tabular}{|l|l|l|l|}
 \hline & $\nu_e$ & $\bar{\nu}_{{e}}$ & $\nu_x,~\bar{\nu}_x$ \\
 \hline \hline $\lambda_{\text{emission}_{\phantom{l}}}^{i}$[km] & $1 / [50 \xi(r)]$ & $1 / [26 \xi(r)]$ & $1 / [10 \xi(r)]$ \\
 \hline $\lambda_{\text {absorb }_{\phantom{l}}}^{i}$[km] & $1 / [50 \xi(r)]$ & $1 / [25 \xi(r)]$ & $1 / [10 \xi(r)]$ \\
 \hline $\lambda_{\text {dir-ch }_{\phantom{l}}}^{i}$[km] & $1 / [50 \xi(r)]$ & $1 / [25 \xi(r)]$ & $1 / [12.5 \xi(r)]$ \\
 \hline
 \end{tabular} 
 \label{Tab:mfp_2}
\end{table}

\begin{table}[t]
\caption{Summary of the type of the $\Theta$-dependent perturbations using the mean free paths in Table~\ref{Tab:mfp_2}.}
\centering
\begin{tabular}{|l||l|}
\hline
Case 2NP & No Perturbations \\ \hline
Case 2HP & $H_{\mathrm{vac}}$ Perturbations  \\ \hline
Case 2CP & $\mathcal{C}_{\mathrm{emission}}$ Perturbations \\ \hline
\end{tabular}
\label{Tab:cases_app}
\end{table}

\begin{figure}
\centerline{Case 2NP (no perturbations)}
\includegraphics[width=0.99\textwidth]{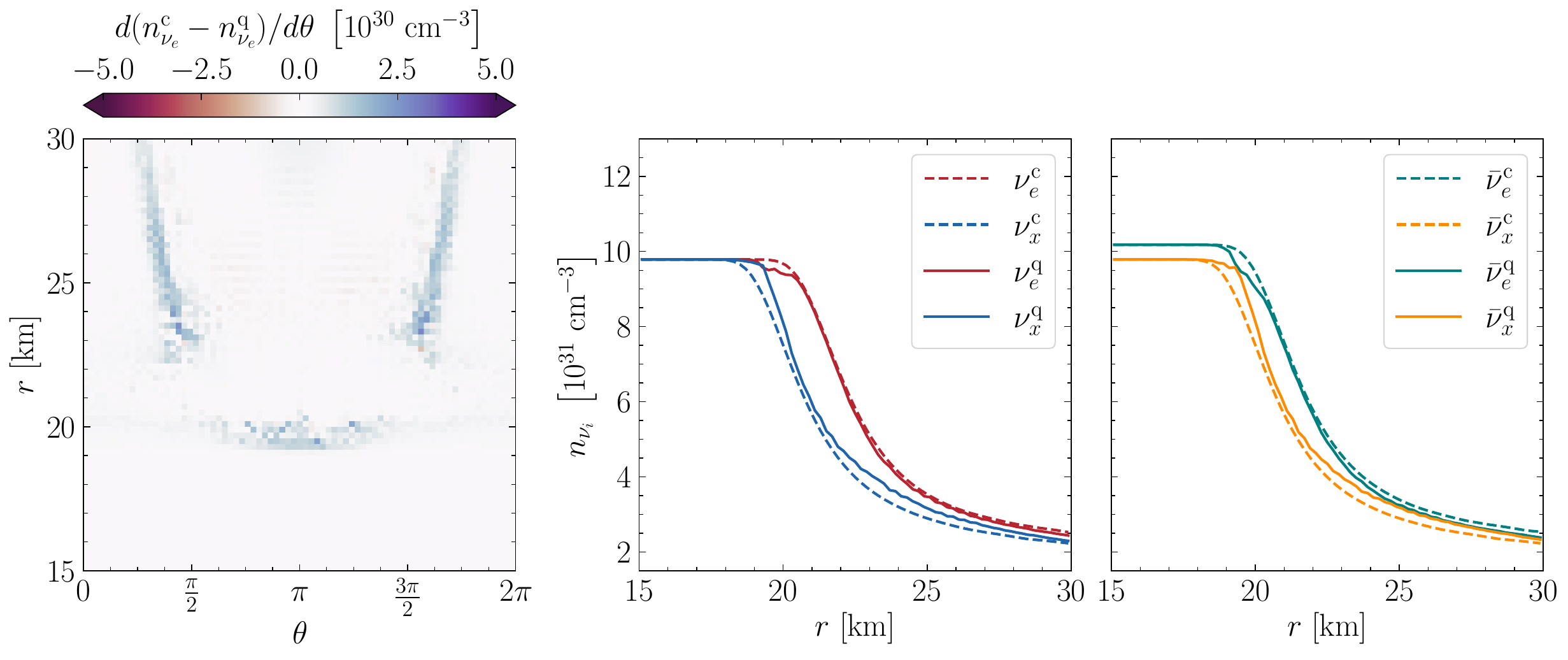}
\caption{Same as Fig.~\ref{Fig:heatmap_angav_case1NP}, but for the collision term modeled as in Table~\ref{Tab:mfp_2} (Case 2NP). The simulation is evolved for $t=50~\mu$s and the plots show an average of the solutions between $t=40~\mu$s and $t=50~\mu$s. The ELN radial evolution allows for negligible flavor conversion between $21$ and $22$~km.
}
\label{Fig:heatmap_angav_case2NP}
\end{figure}

Figure~\ref{Fig:heatmap_angav_case2NP} shows the corresponding quasi-steady state configuration obtained in the absence of perturbations (Case 2NP). Due to the absence of $\Theta$-dependent perturbations, the quasi-steady configuration of Case 2NP corresponds to the one obtained for Case B in Ref.~\cite{Shalgar:2022lvv} (cf.~their Fig.~7) except for the additional term present in the neutrino self-interaction Hamiltonian in our 2D framework and responsible for small variations in the quasi-steady state number densities.
The ELN radial evolution in the classical steady state is such that there are no flavor instabilities developing in a narrow band between $21$ and $22$~km, compared to Case 1NP where flavor instabilities occur at all radii above $20$~km.

Figure~\ref{Fig:heatmap_angav_case2NP} shows the quasi-steady state solution for Case 2HP. Consistently with the findings of Case 1HP, despite the different ELN configuration, we find that perturbations are responsible for small changes in the quasi-steady state configuration (cf.~Fig.~\ref{Fig:heatmap_angav_case2NP} for the case without any perturbations). 
\begin{figure}
\centerline{Case 2HP (perturbations in vacuum Hamiltonian)}
\centering
\includegraphics[width=0.99\textwidth]{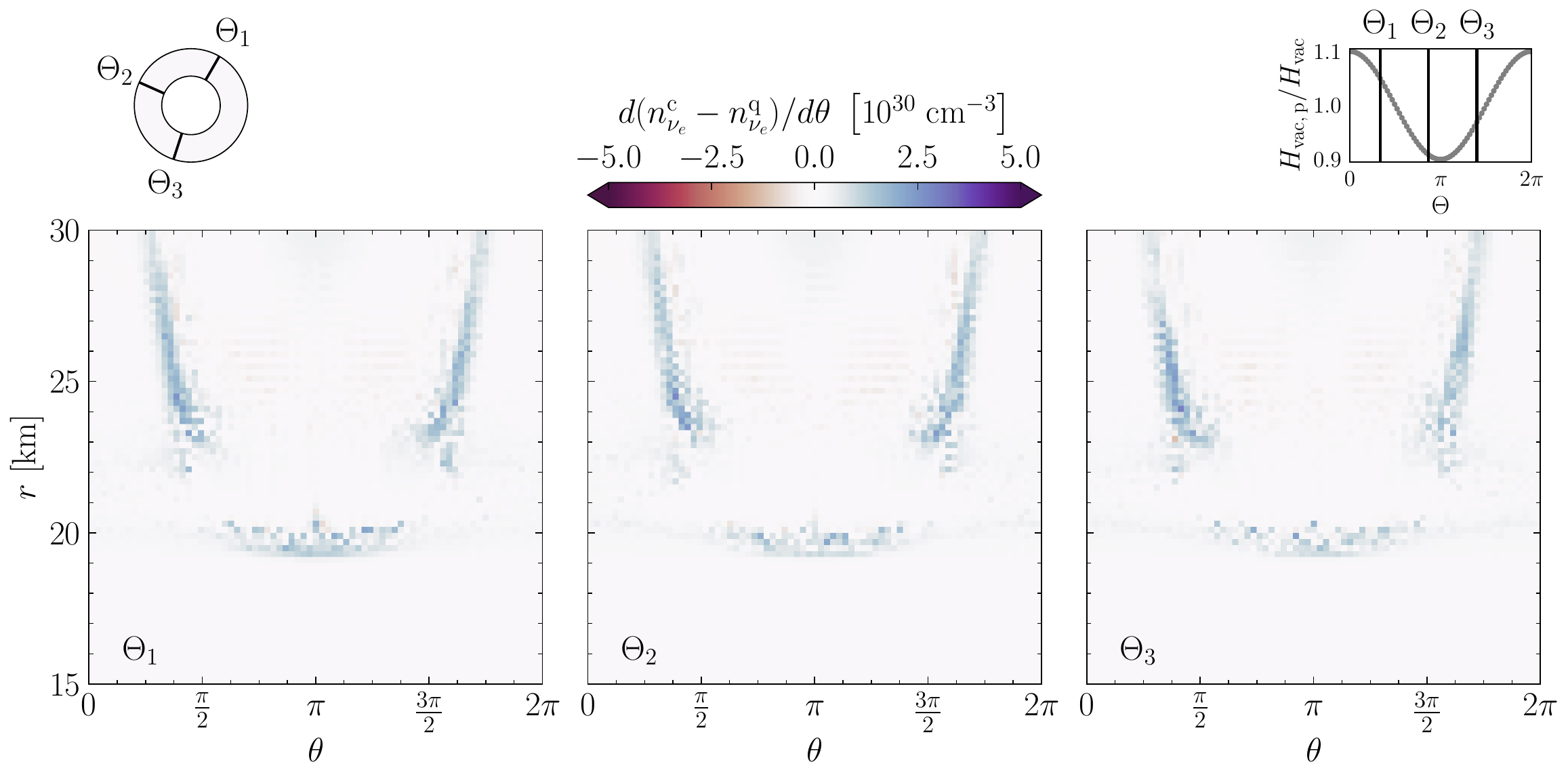}
\includegraphics[width=0.99\textwidth]{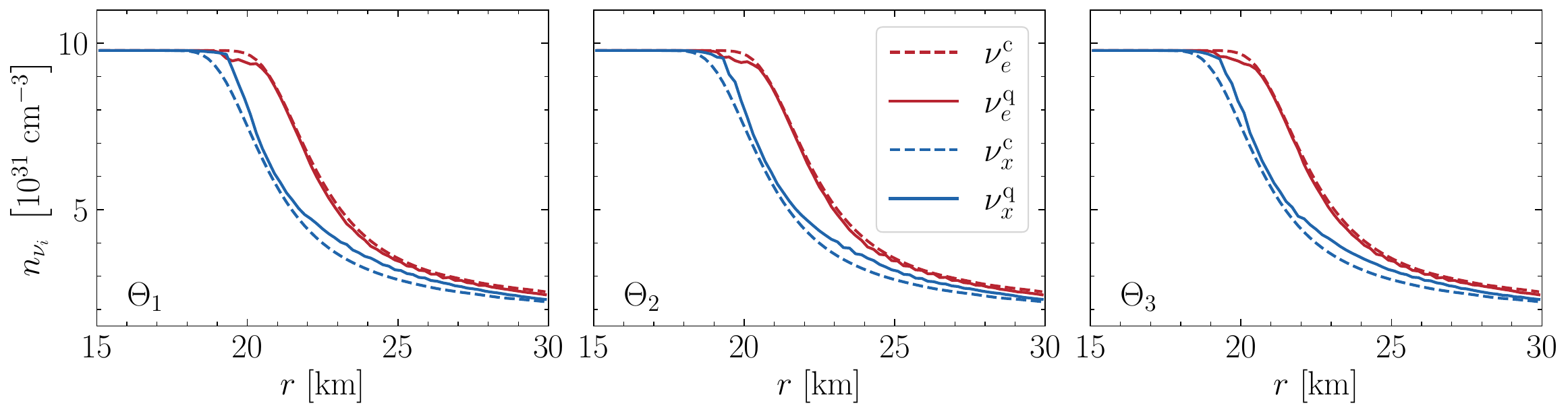}
\caption{Same as Fig.~\ref{Fig:heatmap_angav_tv_case1HP}, but for the collision term modeled as in Table~\ref{Tab:mfp_2} (Case 2HP). The simulation is evolved for $t=50~\mu$s and the plots show an average of the solutions between $t=40~\mu$s and $t=50~\mu$s. Similar to Case 1HP, we find that perturbations in the vacuum term of the Hamiltonian induce small changes in the quasi-steady state configuration.
}
\label{Fig:heatmap_angav_tv_case2HP}
\end{figure}

\begin{figure}
\centerline{Case 2CP (perturbations in collision term)}
\centering
\includegraphics[width=0.99\textwidth]{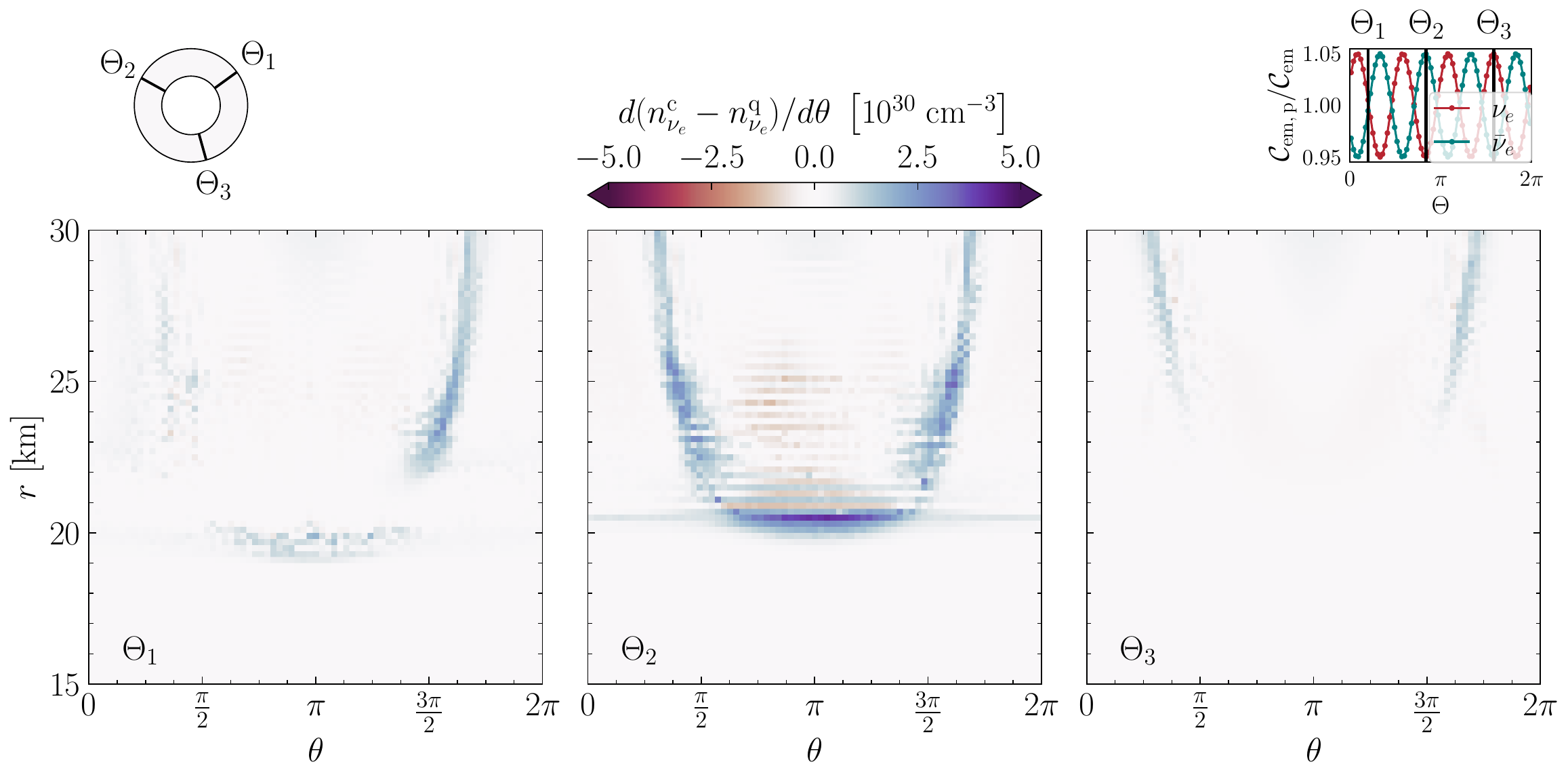}
\includegraphics[width=0.999\textwidth]{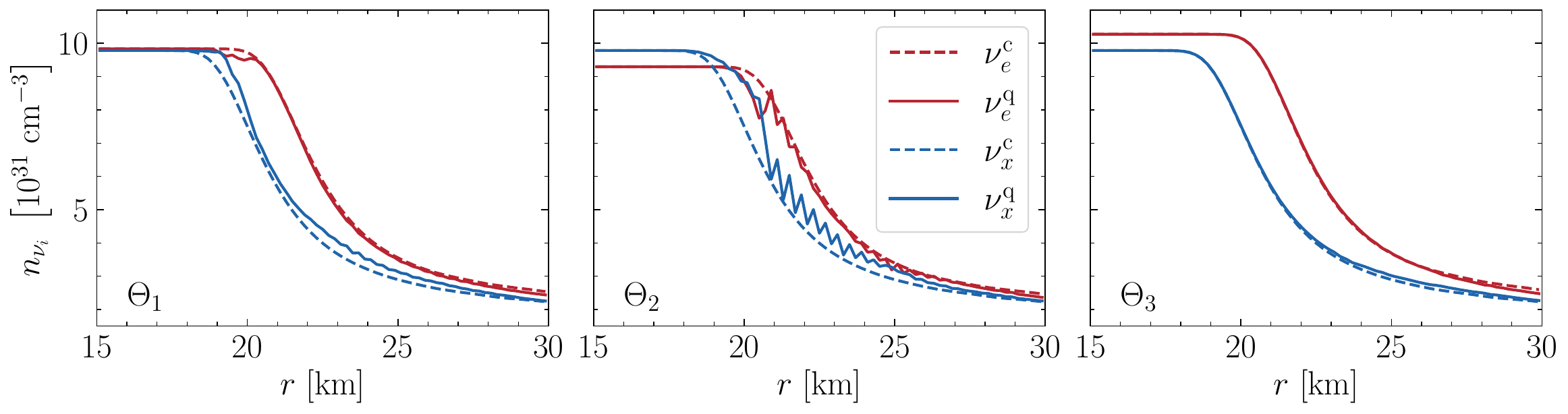}
\caption{Same as Fig.~\ref{Fig:heatmap_angav_ThC_case1CP}, but for the collision term modeled as in Table~\ref{Tab:mfp_2} (Case 2CP). The simulation is evolved for $t=50~\mu$s and the plots show an average of the solutions between $t=40~\mu$s and $t=50~\mu$s. The perturbations in the collision term are responsible for the development of $\Theta$-dependent small-scale structures in the neutrino ensemble.
}
\label{Fig:heatmap_angav_ThC_case2CP}
\end{figure}

\begin{figure}
\centerline{Case 2CP (perturbations in collision term; classical solution)}
\vspace{0.5cm}
\centering
\includegraphics[width=0.999\textwidth]{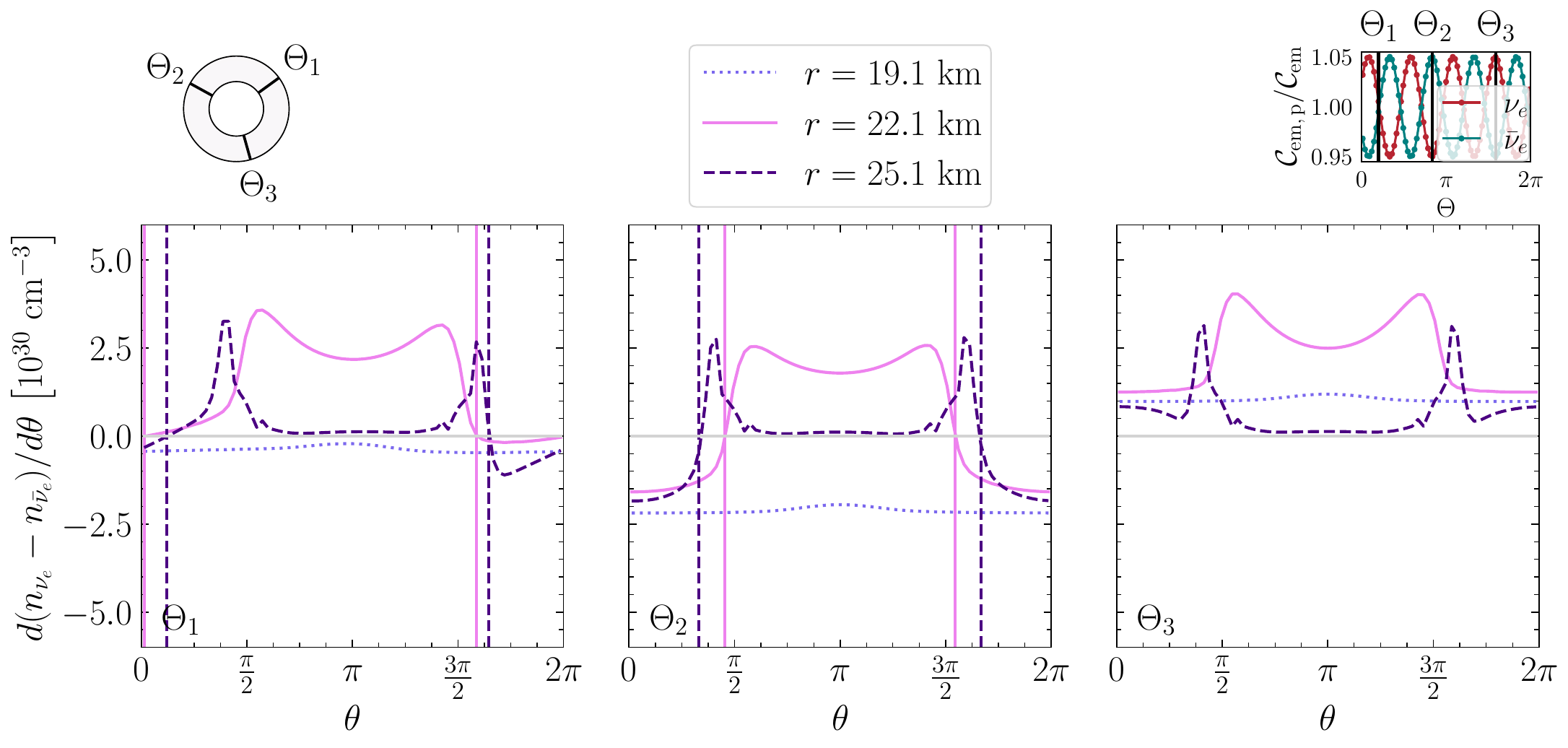}
\caption{Same as Fig.~\ref{Fig:ang_dist_ThC_case1CP}, but for the collision term modeled as in Table~\ref{Tab:mfp_2} (Case 2CP).
}
\label{Fig:ang_dist_ThC_case2CP}
\end{figure}

Figure~\ref{Fig:heatmap_angav_ThC_case2CP} is the analog of Fig.~\ref{Fig:heatmap_angav_ThC_case1CP}, but it has been obtained for the collision term modeled as in Table~\ref{Tab:mfp_2} (Case 2CP). Similar to Case 1CP, we find that the perturbations in the collision term introduce $\Theta$-dependent fluctuations in the $\nu_e$ and $\bar\nu_e$ local distribution densities with $\Theta$-dependent implications on the flavor conversion physics. 
It is interesting to note that, in Case 2CP shown (Fig.~\ref{Fig:heatmap_angav_ThC_case2CP}), the regions of flavor conversion do not match the peaks in the $\nu_e$ emission, rather the ones in the $\bar\nu_e$ emission. This can be explained by inspecting Fig.~\ref{Fig:ang_dist_ThC_case2CP} which shows the angular distributions: in the locations with more $\bar\nu_e$, the ELN crossings form at two angles for larger radii ($\Theta_2$), and in the sites with more $\nu_e$'s, the ELN distribution does not show any crossings at the selected radii ($\Theta_3$). Furthermore, for $r=25.1$~km at $\Theta_3$ there are no ELN crossings, but when comparing to the contour plot of $\Theta_3$ in Fig.~\ref{Fig:heatmap_angav_ThC_case2CP}, flavor conversion has developed at $r=25.1$~km, suggesting that neutrino advection contributes to the spreading of flavor waves.

\bibliographystyle{JHEP}
\bibliography{references}

\end{document}